\RequirePackage{etex}
\RequirePackage{easybmat}

\documentclass[11pt,a4paper,twoside]{article}
\usepackage{fullpage}
\usepackage[bookmarks=true,hyperfigures=true]{hyperref}
\usepackage[final]{graphicx}
\usepackage[T1]{fontenc}

 \usepackage{pstricks-add}

\usepackage{amsmath,amssymb}
\usepackage{enumerate}  
\usepackage[stdtext]{youngtab}
\usepackage{young}

\numberwithin{equation}{section}

\newcommand{\alg}[1]{\mathfrak{#1}}
\newcommand{\Qop}{{\rm \bf{Q}}}
\newcommand{\Xop}{{\rm \bf{X}}}
\newcommand{\Top}{{\rm \bf{T}}}

\newcommand{\Gbb}{{\mathbb G}}
\newcommand{\Xbf}{{\mathbf X}}
\newcommand{\osca}{\mathbf{a}}
\newcommand{\oscb}{\mathbf{b}}
\newcommand{\oscc}{\mathbf{c}}
\newcommand{\osch}{\mathbf{h}}
\newcommand{\oscbp}{{\mathbf b}^{\dagger}}
\newcommand{\oscbm}{{\mathbf b}^{\phantom{\dagger}}}

\newcommand{\tr}{{\rm Tr}} 
 
\newcommand{\specbaz}{z}
\newcommand{\inv}{\mbox{Inv}}

\newcommand{\Rff}{{\bf R}}
\newcommand{\Rfi}{{\bf L}_I}
\newcommand{\Rfl}{{\mathcal{L}}}
\newcommand{\Rli}{\mathcal{R}_{ I}}

\newcommand{\Rlam}{\mathfrak{R}_{}}

\newcommand{\beq}{\begin{equation}}
\newcommand{\eeq}{\end{equation}}
\newcommand{\sfrac}[2]{{\textstyle\frac{#1}{#2}}}
\newcommand{\half}{\sfrac{1}{2}}

\begin{document}

%%%%%%%%%%%%%%%%%%%%%% TITLE PAGE %%%%%%%%%%%%%%%%%%%%%%%%%%%%%%%%%%%
\thispagestyle{empty}
\begin{flushright}\footnotesize
\texttt{HU-Mathematik:~2011-22}\\
\texttt{HU-EP-11/52}\\
\texttt{AEI-2011-081}\\
\vspace{0.5cm}
\end{flushright}
\setcounter{footnote}{0}

\begin{center}
{\Large\textbf{\mathversion{bold}
Baxter Operators and Hamiltonians for ``nearly all'' Integrable Closed $\alg{gl}(n)$ Spin Chains
}\par}
\vspace{15mm}

{\sc Rouven Frassek $^{a,b}$, Tomasz {\L}ukowski $^{a}$, Carlo Meneghelli $^{a,b,c,d}$,\\
Matthias Staudacher $^{a,b}$}\\[5mm]

{\it $^a$ Institut f\"ur Mathematik und Institut f\"ur Physik, Humboldt-Universit\"at zu Berlin\\
Johann von Neumann-Haus, Rudower Chaussee 25, 12489 Berlin, Germany
}\\[5mm]

{\it $^b$ Max-Planck-Institut f\"ur Gravitationsphysik, Albert-Einstein-Institut\\
    Am M\"uhlenberg 1, 14476 Potsdam, Germany
}\\[5mm]
    
{\it $^c$ Fachbereich Mathematik, Universit\"at Hamburg\\
Bundesstra{\ss}e 55, 20146 Hamburg, Germany
}\\[5mm]

{\it $^d$ Theory Group, DESY,\\
Notkestra{\ss}e 85, D-22603, Hamburg, Germany
}\\[5mm]    

\texttt{rfrassek@physik.hu-berlin.de}\\
\texttt{lukowski@mathematik.hu-berlin.de}\\
\texttt{carlo.meneghelli@gmail.com}\\
\texttt{matthias@aei.mpg.de}\\[52mm]

\textbf{Abstract}\\[2mm]
\end{center}
We continue our systematic construction of Baxter Q-operators for spin chains, which is based on certain degenerate solutions of the Yang-Baxter equation. Here we generalize our approach from the fundamental representation of  $\alg{gl}(n)$ to generic finite-dimensional representations in quantum space. The results equally apply to non-compact representations of highest or lowest weight type. We furthermore fill an apparent gap in the literature, and provide the nearest-neighbor Hamiltonians of the spin chains in question for all cases where the $\alg{gl}(n)$ representations are described by  rectangular Young diagrams, as well as for their infinite-dimensional generalizations. They take the form of digamma functions depending on operator-valued shifted weights.

%%%%%%%%%%%%%%%%%%%%%%%%% END OF TITLE PAGE %%%%%%%%%%%%%%%%%%%%%%%%%%%%%

\newpage

%%%%%%%%%%%%%%%%%%%%%%%%%%%%%%%%%%%%%%%%%%%%%%%%
%%%%%%%%%%%%%%%%%%%%%%%%%%%%%%%%%%%%%%%%%%%%%%%%%%%%%
\section{Introduction, Overview and Outlook}
%%%%%%%%%%%%%%%%%%%%%%%%%%%%%%%%%%%%%%%%%%%%%%%%
%%%%%%%%%%%%%%%%%%%%%%%%%%%%%%%%%%%%%%%%%%%%%%%%%%%%%
We continue our systematic approach to the construction of Baxter Q-operators for those types of quantum spin chains that are based on rational R-matrices \cite{Bazhanov:2010ts}, \cite{Bazhanov:2010jq}, \cite{Frassek:2010ga}. Our program is motivated by the desire to gain a deeper understanding of the integrable structure of the free/planar AdS/CFT system, where spin chains appear in the weak coupling limit. We will, however, not directly explore this connection in this paper, and refer the reader to  the remarks made in \cite{Bazhanov:2010ts}, \cite{Frassek:2010ga} as well as to the overview article series \cite{Beisert:2010jr} for further information. Instead, we will proceed to extend the methodology of \cite{Bazhanov:2010jq}, where only the fundamental representation was considered, to the case where the state space of the quantum spin chain is allowed to be a tensor product of arbitrary finite-dimensional or non-compact $\alg{gl}(n)$ representations of highest or lowest weight type. Note that non-compact lowest-weight representations are precisely the ones which are needed in the AdS/CFT spectral problem, albeit in a supersymmetric generalization. The latter will be the subject of future work. 

In this article we restrict ourselves to the purely ``bosonic'' $\alg{gl}(n)$ case, and provide in Section~\ref{sec:qoperators} a very general construction of the Baxter Q-operator. In Section~\ref{sec:review} we briefly recall our earlier work, where we had algebraically found a new class of Lax operators for the fundamental representation of $\alg{gl}(n)$ by studying certain singular solutions of the Yang-Baxter equation. We also set up the generalization to general representations in quantum space. Section~\ref{sec:oscpart} proceeds to find the needed, more general solutions of the Yang-Baxter equation, yielding novel (quantum) Lax operators for arbitrary $\alg{gl}(n)$ representations. Of special importance in our construction is the non-commutative Cayley-Hamilton theorem, which is reviewed in Appendix \ref{app:Reving}. We then succinctly discuss the fusion properties of the Lax operators in Section~\ref{sec:fusionrev}. Up to this point our work is purely algebraic, and not restricted to specific representations. After discussing in \ref{sec:analytic} certain analytic properties of the derived new class of Lax operators, we proceed, as in our earlier work, in Section \ref{sec:Qconstruction} to construct Baxter operators by first building suitable monodromies based on the new solutions, and subsequently tracing over the oscillator representations of the auxiliary space, a technique pioneered in \cite{Bazhanov:1998dq}. At this point we choose to restrict ourselves to lowest or highest weight representations, deferring the most general case to future work. As shown in Section~\ref{sec:exappl}, functional relations follow in the usual way from the fusion relations, and systems of Bethe equations are then deduced in a straightforward manner from these relations. We thereby rederive the most general system of nested $\alg{gl}(n)$ Bethe equations, first obtained in \cite{Kulish:1983rd} by means of the Bethe ansatz (which our method entirely avoids). A closely related method to obtain these equations called ``analytical Bethe ansatz'' was presented in \cite{Arnaudon:2004vd}. See also \cite{Kazakov:2007fy}, where this generic system of Bethe equations was also derived (including the supersymmetric case) from the fusion relations of the eigenvalues of Baxter operators, without providing, however, an operatorial construction. Our construction has the additional advantage that whenever Q-operators are diagonalizable then the set of solutions of the Bethe equations is complete. To conclude our Q-operator construction, we show in Section~\ref{sec:qopinuse} and in particular in Appendix~\ref{app:Qcode} that our method is also of very practical value, if one e.g.~is interested in explicitly constructing Q-operators for chains of small length. This is done by providing {\it Mathematica}\texttrademark\, programming code.

Section~\ref{sec:hamiltonianchapter}, which may be read largely independently, is concerned with providing, tongue-in-cheek, interesting problems to the solutions found in Section~\ref{sec:qoperators}. Baxter operators and transfer matrices for integrable spin chains form a large family of commuting operators. For applications, one is often interested in a special member of this family: a local, nearest-neighbor {\it Hamiltonian}. Integrable Hamiltonians for general $\alg{gl}(2)$  representations have been known for a long time, \cite{Kulish:1981gi}, \cite{Babujian:1982ib}, \cite{Babujian:1983ae}. Their derivation is based on the general method of finding a suitable transfer matrix with a special ``regular point'' in the spectral parameter plane, where the matrix becomes the shift operator on the chain \cite{Tarasov:1983cj}. The Lax operator (R-matrix) used in the construction of these special transfer matrices intertwines two identical representations of the type present in the quantum space of the spin chain. For $\alg{gl}(2)$ it takes the form of a ratio of Euler gamma functions,  whose arguments depend on the joint quadratic Casimir operator of the intertwined representations. The Hamiltonian density of the spin chain is then obtained as the logarithmic derivative of the R-matrix w.r.t.~the spectral parameter at its shift point value, and turns into (a sum of) digamma functions. See in particular \cite{Faddeev:1996iy} for a beautiful discussion of this approach. To our knowledge, an explicit generalization of these $\alg{gl}(2)$ formulas for the Lax operator and associated nearest-neighbor Hamiltonian density to the case of general $\alg{gl}(n)$ spin chains is lacking in the literature. The state of the art appears to be the so-called ``tensor product graph method'' \cite{MacKay:1991bj}, \cite{Gould:2002de} (and further references therein) for multiplicity-free representations, which expresses the R-matrices as a (finite or infinite) series involving projectors on all irreducible representations occurring in the two-site tensor product decomposition. We should also mention two other approaches to the construction of R-matrices. For finite-dimensional representations of $\mathfrak{gl}(n)$, one can also obtain R-matrices by applying the standard fusion procedure to Yang's R-matrix, see e.~g.~\cite{Nazarov}. Furthermore, for continuous series representations of $\mathfrak{gl}(n)$ R-matrices are given in a remarkable factorized form in \cite{Derkachov:2010zz}. The earlier constructions \cite{Kulish:1983rd},  \cite{Arnaudon:2004vd} do not propose the $\alg{gl}(n)$ generalization of the explicit $\alg{gl}(2)$ Hamiltonians. This prompted us to (partially) fill this apparent gap in the literature, and to derive Hamiltonians for all those $\alg{gl}(n)$ highest or lowest weight representations that satisfy a ``generalized rectangularity condition'' (this includes the cases where the two-site tensor product is multiplicity-free). We set up the derivation in Section \ref{sec:IntroHam}, and then solve in Section~\ref{sec:Rlam} the Yang-Baxter equation for R-matrices intertwining two such rectangular representations. Technically, the method very closely follows the one employed in Section~\ref{sec:qoperators}. The result is expressed as products of gamma functions, whose arguments are linear expressions involving the spectral parameter and operator valued {\it shifted weights}. This explicit result is checked against the formal expression known from the ``tensor product graph method'' in Appendix~\ref{app:TPG}.   In Section~\ref{sec:hamiltonian} we find the shift point of these R-matrices, where they turn into the permutation operator. Taking a logarithmic derivative, we find the Hamiltonian densities of quantum spaces in arbitrary rectangular representations as digamma functions functionally depending on the shifted weights. Some details are delegated to Appendix \ref{app:ham}. In Section~\ref{sec:dispersion} we find, using the functional relations, the dispersion laws allowing to extract the spectrum of the Hamiltonians from the solution to the system of Bethe equations derived in Section~\ref{sec:exappl}. This is in line with our above statement that we provide a natural diagonalization {\it problem} (local Hamiltonians and their dispersion laws) to the {\it solution} (the systems of Bethe equations)!
%
% \newline
%\indent 

Let us mention some other recent work on the Baxter operators of $\alg{gl}(n)$ spin chains. An alternative method employing the so-called co-derivative was proposed in \cite{Kazakov:2010iu}. It would be interesting to see whether it may be generalized from the fundamental representation to general highest/lowest weight representations as in our current work. Furthermore, there are the articles \cite{Chicherin:2011sm},\cite{Chicherin:2011nj} (see also many references therein for the general method), where the case of compact and non-compact $\alg{gl}(2)$ representations is treated by an alternative procedure. It would be interesting to see whether this approach allows for an equally unified description of compact and lowest/highest non-compact $\alg{gl}(n)$ representations, as is needed when applying the Q-operator construction to e.g.~the AdS/CFT system.

We have obtained all of our new Lax operators for Q-operators (Section \ref{sec:qoperators}) and for the transfer matrices with regularity property for spin chains in general representations (Section~\ref{sec:hamiltonianchapter}) by explicitly solving the YBE on a case-by-case basis. It would be very interesting and, perhaps, conceptually more satisfying to recover all these results from Drinfel'd's universal R-matrix, see e.~g.~\cite{Boos2010}, \cite{Ridout:2011wx} for some related, recent work in this direction. This, as well as the generalization to the supersymmetric case, is left to future investigation.

%%%%%%%%%%%%%%%%%%%%%%%%%%%%%%%%%%%%%%%%%%%%%%%%
%%%%%%%%%%%%%%%%%%%%%%%%%%%%%%%%%%%%%%%%%%%%%%%%%%%%%
\section{Q-operators}
\label{sec:qoperators}

\subsection{Review and Outset}
\label{sec:review}

The quantum mechanical state space $\mathcal{V}$ of an integrable, homogeneous lattice chain model usually takes the form 
\begin{equation}
\mathcal{V}=\underbrace{V\otimes V\otimes \ldots \otimes V}_{L-\mbox{\small times}}
\label{quantumspace}
\end{equation}
of an $L$-fold tensor product of equal ``local'' representation spaces $V$ of some algebra. Here $L$ is the length of the chain, and $\mathcal{V}$ is frequently abbreviated to ``quantum space''. In this article we take the algebra acting on $V$ to be the Lie algebra $\alg{gl}(n)$, and will, as is customary, frequently refer to the lattice model as a ``spin'' chain, even though this name strictly speaking only applies to the case $n=2$. 

Historically, integrable spin chains were discovered through solving, by means of Bethe's famous ansatz, the diagonalization problem of some given, specific, nearest-neighbor Hamiltonians acting on $\mathcal{V}$. The {\it quantum inverse scattering method} (QISM), see \cite{Faddeev:1996iy} for a very authoritative review, instead put the emphasis on the rather systematic construction of a commuting family of transfer matrix operators, which may be constructed by starting from the Yang-Baxter equation (YBE), and depend on a free diagonalization-independent parameter, the {\it spectral parameter}. The integrable nearest-neighbor Hamiltonian is then derived from this systematic approach by expanding the ``correct'' transfer matrix operator around the ``correct'' value of the spectral parameter. See also Section~\ref{sec:hamiltonianchapter} for a more detailed discussion.

Curiously, while the usual transfer matrix operators forming the commuting family depend on the spectral parameter and an additional label, the representation in so-called auxiliary space \cite{Faddeev:1996iy}, the family is even bigger, and also encompasses certain {\it singular} transfer matrices termed Q-operators. Here ``singular'' carries a double sense: for one, according to the usual QISM these operators diverge unless the global $\alg{gl}(n)$ symmetry of $\mathcal{V}$, which for periodic spin chain boundary conditions is inherited from the  $\alg{gl}(n)$-invariance of $V$, is broken by certain regulating ``angles''. For another, these operators are constructed from certain singular solutions of the YBE. All this is discussed in our earlier work \cite{Bazhanov:2010jq}, where we constructed Q-operators for the integrable spin chain with $V$ being in the fundamental representation of  $\alg{gl}(n)$. In the following we extend our analysis to the general case where $V$ is the representation space of any $\mathfrak{gl}(n)$ irreducible representation.

According to the idea presented in \cite{Bazhanov:2010jq}  we construct Q-operators as special transfer matrices, where the auxiliary space is taken to be an appropriately chosen Yangian algebra representation. Before proceeding to the determination of the relevant R-operator we briefly review the representations of the Yangian algebra introduced in \cite{Bazhanov:2010jq}. The YBE
\begin{equation}\label{RLL}
\Rff(z_1-z_2)(\mathbb{L}(z_1)\otimes 1)(1\otimes \mathbb{L}(z_2))=(1\otimes \mathbb{L}(z_2))(\mathbb{L}(z_1)\otimes 1)\Rff(z_1-z_2)
\end{equation}
with
\begin{equation}\label{solfund}
\Rff(z)=z\,\mathbf{I}+\mathbf{P}
\end{equation}
provides the defining relation for the Yangian algebra $\mathcal{Y}=Y(\mathfrak{gl}(n))$. The $\mathbb{L}$-operators of the form\footnote{We are changing notation with respect to \cite{Bazhanov:2010jq} as follows:  \begin{equation*}
 \oscbp_{a\dot{b}}\rightarrow \bar{\osca}^{\dot{b}}_a\,,\qquad  \qquad \qquad \oscbm_{\dot{a} b} \rightarrow \osca^b_{\dot{a}}\,.
\end{equation*}}
\begin{equation}\label{Rfi}
\Rfi(z)=\left(
\begin{BMAT}(@,30pt,30pt)
{c.c}{c.c}
z\,\delta^{a}_{b}+H^{a}_{b}&\bar{\osca}^{\dot a}_{b}\\
-\osca^{a}_{\dot b}&\delta^{\dot a}_{\dot b}
\end{BMAT}
\right),
\end{equation}
with 
\begin{equation}
\label{smallj}
 H^a_b=\bar{\mathcal{J}}^a_b-\sum_{\dot c\in \bar I}\bar{\osca}_b^{\dot{c}}\,\osca^a_{\dot{c}}-\frac{|\bar{I}|}{2}\,\delta^a_b\,,
\end{equation}
satisfy the relation \eqref{RLL} with $\mathbb{L}(z)=\Rfi(z)$ provided that $\bar{\mathcal{J}}^{a}_{b}$ form a $\mathfrak{gl}(I)$ subalgebra and $(\bar\osca^{\dot a}_{b},\osca^{b}_{\dot a})$ are families of oscillators, namely
\begin{eqnarray}\label{oscillators}
&&\left[ \osca_{a}^{b},\bar\osca_{c}^{d}\right] = \delta_{a}^{d}\delta_{c}^{b}\,,\\
 &&\left[ \bar{\mathcal{J}}^a_b,\bar{\mathcal{J}}^c_d\right]=\delta^a_d\,\bar{\mathcal{J}}^c_b-\delta^c_b\,\bar{\mathcal{J}}^a_d\,,\\
 &&\left[ \bar{\mathcal{J}}_{a}^{b},\osca_{c}^{d}\right]=0= \left[ \bar{\mathcal{J}}_{a}^{b},\bar\osca_{c}^{d}\right]\,.
\end{eqnarray}
 Here $I$ is any subset of the index set $\{ 1,\ldots,n\}$, and $\bar I$ is its complementary set. For graphical convenience the $\mathbb{L}$-operator in \eqref{Rfi} is presented in the case where the set $I=\{1,\ldots,|I|\}$. Any other $\Rfi(z)$ can be obtained by appropriate permutation of rows and columns. We use the notation $a,b, \ldots$ for indices $a,b, \ldots\in I$ and $\dot a,\dot b\,\ldots$ for $\dot a,\dot b\,\ldots \in \bar I$. The solutions \eqref{Rfi} provide an evaluation homomorphism of the infinite-dimensional Yangian algebra into some finite dimensional algebra.
The latter we denote by
\begin{equation}
\label{AIchapt3}
\mathcal{A}_{I}=\alg{gl}(I)\otimes \mathcal{H}^{(I,\bar{I})}\,.
\end{equation}
There are $2^n$ homomorphisms of such form corresponding to all possible choices of the set $I$ and every representation of $\mathfrak{gl}(I)$ in $\mathcal{A}_{I}$ provides a distinguished representation of the Yangian algebra. When the set $I $ is chosen to be $I=\{ 1,\ldots,n\}$ we find the well-known result
\begin{equation}\label{Rfl}
\Rfl(z)=\mathbf{L}_{\{1,\ldots,n\}}(z)=z\,\mathbf{I}+e_{b}^{a}\otimes J_{a}^{b}\,,
\end{equation}
where by $e^{a}_{b}$ we denote $\mathfrak{gl}(n)$ generators in the fundamental representation and $J_{b}^{a}$ generate the $\mathfrak{gl}(n)$ algebra
\begin{equation}
\left[J^{a}_{b},J_{d}^{c}\right]=\delta_{b}^{c}\,J_{d}^{a}-\delta_{d}^{a}\,J_{b}^{c}\,.
\end{equation}

While the Lax operators $\Rfi(z)$ are sufficient to construct Q-operators for the fundamental representation, the first step in the generalization of \cite{Bazhanov:2010jq} is to find ``R-operators for Q-operators''. These operators are defined as intertwiners between a given representation of the algebra $\mathcal{A}_I$ and the chosen representation of the $\mathfrak{gl}(n)$ algebra related to $V$. However, it is convenient not to specify these representations and work with the abstract generators of $\mathcal{A}_I$ in \eqref{Rfi} and $\mathfrak{gl}(n)$ in \eqref{Rfl}. The defining relation for R-operators is then the YBE of the form
\begin{equation}\label{YBE.expl}
\Rfl(z_1)\Rfi(z_2)\Rli(z_2-z_1)=\Rli(z_2-z_1)\Rfi(z_2)\Rfl(z_1)\,.
\end{equation}
The intertwiners $\mathcal{R}_I(z)$ are the basic building blocks of the transfer matrices of the form
\begin{equation}\label{operatorastrace}
\tr_{I} \left[ \mathcal{D}_{I}\,\mathcal{R}_{I}(z)\otimes \mathcal{R}_{I}(z)\otimes \ldots \otimes \mathcal{R}_{I}(z)\right].
\end{equation}
The precise form of the regulator $\mathcal{D}_I$ and the definition of trace will be explained in the following.

%%%%%%%%%%%%%%%%%%%%%%%%%%%%%%%%%%%%%%%%%%%%%%%%%%%%%%%%%%%%
\subsection{New Solutions of the Yang-Baxter Equation}
\label{sec:oscpart}
%%%%%%%%%%%%%%%%%%%%%%%%%%%%%%%%%%%%%%%%%%%%%%%%%%%%%%%%%%%%
This section contains the derivation of the intertwiner $\mathcal{R}_I(z)$ entering \eqref{YBE.expl}. 
We will use some important results of classical invariant theory.
The latter appear to be closely connected with the theory of Yangians, see \cite{molevbook} for more details 
and references to the large literature on the subject.
In order to solve \eqref{YBE.expl} for $\Rli(z)$ it is convenient to make use of the difference property of R-operators. Using the explicit form of the solutions \eqref{Rfi} and \eqref{Rfl} we obtain the commutation relations

\begin{equation}
\label{iterwiningFULLI}
[\osca^b_{\dot{a}},\Rli]=J^b_{\dot{a}}\, \Rli\,,\qquad [\bar{\osca}^{\dot{a}}_b,\Rli]=\Rli\,J^{\dot{a}}_b\,, \qquad [\Rli,J^a_b+ H^a_b]=0\,.
\end{equation}
Without loss of generality we may rewrite $\Rli(z)$ in the factorized form
\beq
\label{R0tobe}
\Rli(z)\,=\,e^{\bar{\osca}^{\dot{c}}_c\,J^c_{\dot{c}}}\,\,\mathcal{R}_{0,I}(z)\,\,e^{-\osca^c_{\dot{c}}\,J^{\dot{c}}_c}\,.
\eeq
Remarkably, \eqref{iterwiningFULLI} immediately implies that $\mathcal{R}_{0,I}(z)$ does not depend on oscillators and that it is $\mathfrak{gl}(I)$ invariant. 
In addition, using \eqref{R0tobe}, we obtain $n\times n$ equations for $\mathcal{R}_{0,I}(z)$ from the YBE, naturally organized into four blocks:

\begin{equation}
\label{eqcase1.2}
 [\mathcal{R}_{0,I}(z),J^{\dot{a}}_{\dot{b}}]=0\,,
\end{equation}
\begin{equation}
\label{eqIlookat}
 \mathcal{R}_{0,I}(z)\left(( z-\sfrac{|\bar{I}|}{2}) \, J^{\dot{a}}_{b}+\bar{\mathcal{J}}^{c}_{b}J^{\dot a}_{c} -J^{\dot{c}}_{b}\,J^{\dot{a}}_{\dot{c}}\right)\,= J^{\dot{a}}_{b}\,\mathcal{R}_{0,I}(z)\,,
\end{equation}
\begin{equation}
\label{eqIlookat2}
 \mathcal{R}_{0,I}(z)J_{\dot{a}}^{b}=\left( (z-\sfrac{|\bar{I}|}{2}) \, J_{\dot{a}}^{b}+\bar{\mathcal{J}}^{b}_{c}J^{c}_{\dot a}-J^{\dot{c}}_{\dot{a}}\,J^{b}_{\dot{c}}\right)\, \mathcal{R}_{0,I}(z)\,,
\end{equation}
\begin{equation}
\label{otherbef}
 \mathcal{R}_{0,I}(z)\left(( z-\sfrac{|\bar{I}|}{2}) \, J^{a}_{b}+\bar{\mathcal{J}}^{d}_{b}J^{a}_{d}-J^{\dot{c}}_{b}\,J^{a}_{\dot{c}}\right)=
\left(( z-\sfrac{|\bar{I}|}{2}) \,J^{a}_{b} +J^{d}_{b}\bar{\mathcal{J}}^{a}_{d}-J^{\dot{c}}_{b}\,J^{a}_{\dot{c}}\right)\, \mathcal{R}_{0,I}(z)\,.
\end{equation}
Let us focus on the case relevant for the construction of the Q-operators. This means that we restrict to, cf.~\eqref{AIchapt3},
\begin{equation}\label{minimalrepresentation}
\tilde{\mathcal{A}}_{I}\equiv \cdot\otimes \mathcal{H}^{(I,\bar{I})}\,,
\end{equation}
where we take the singlet representation, denoted by the dot $\cdot$, of the subalgebra $\mathfrak{gl}(I)$ - we call \eqref{minimalrepresentation} a {\em minimal representation}. This means that the generators $\bar{\mathcal{J}}$ in \eqref{smallj} act trivially on (i.e.~annihilate) all states of a given minimal representation. We will nevertheless keep equations \eqref{eqcase1.2}-\eqref{otherbef} in their full form because we will need them to prove the fusion relations in the following section. 
In the case of minimal representations \eqref{otherbef} follows directly from \eqref{eqIlookat} and \eqref{eqIlookat2}, and the solution of \eqref{eqIlookat} satisfies also \eqref{eqIlookat2}. In the following we focus on \eqref{eqIlookat} under the restriction that $\mathcal{R}_{0,I}(z)$ can be expressed as a function of the $\mathfrak{gl}(\bar I)$ Casimir operators\footnote{We believe that this condition follows from 
\begin{equation*}
 [\mathcal{R}_{0,I}\,,J^a_b\,]=0\,,\qquad [\mathcal{R}_{0,I}\,,J^{\dot{a}}_{\dot{b}}\,]=0\,,\qquad [\mathcal{R}_{0,I}\,,J^{\dot{c}}_{b}\,J^{a}_{\dot{c}}\,]=0\,,
\end{equation*}
which are specializations, respectively, of the last equation in \eqref{iterwiningFULLI}, \eqref{eqcase1.2} and \eqref{otherbef} in the case of minimal representations. Clearly $\mathcal{R}_{0,I}$ can be considered as a function of the Casimir operators of $\mathfrak{gl}(n)$ as well. These are just constants in a given irreducible representation and will not enter the discussion regarding the determination of $\mathcal{R}_{0,I}$.
}. 

As a warm-up exercise, and in order to investigate what types of solutions may be expected, we solve \eqref{eqIlookat} for the simple case when $|\bar{I} |=1$.  As the dotted indices take only one value we can simplify it to 
\begin{equation}\label{Ibaris1}
 \mathcal{R}_{0,I}(z)\,\left(z+\half-J^{\dot a}_{\dot a} \right)\,J^{\dot a}_b=J^{\dot a}_b\,\mathcal{R}_{0,I}(z).
\end{equation}
In this case $\mathcal{R}_{0,I}(z)$ can be expressed as a function of the linear Casimir $J_{\dot a}^{\dot a}$ of $\mathfrak{gl}(\bar I)$. Now, using 
\begin{equation}
 [J^{\dot a}_{\dot a},J^{\dot a}_b]=J^{\dot a}_b\,,\qquad \qquad\Rightarrow \qquad\qquad J^{\dot a}_b\,f(J^{\dot a}_{\dot a})=f(J^{\dot a}_{\dot a}-1)\,J^{\dot a}_b\,,
\end{equation}
we can turn \eqref{Ibaris1} into a simple system of first-order recurrence relations.
These relations may be solved up to a spectral parameter-dependent periodic function in $J^{\dot a}_{\dot a}$ of period 1. The solution of \eqref{Ibaris1} can thus be written with the help of the Euler gamma function as
\begin{equation}\label{ibaris1sol}
\mathcal{R}_{0,\overline{\{\dot{a}\}}}(z)=\rho(z,J^{\dot a}_{\dot a})\,\Gamma(z+\half-J_{\dot a}^{\dot a})\,.
\end{equation}

Next, we observe that a similar factorized form of \eqref{Ibaris1} holds also in case the generators of the algebra may be written in Schwinger form as $J_{B}^{A}=\bar\osca^A \osca_B$ or $J_{B}^{A}=\bar\oscc^A \oscc_B$, where, respectively, bosonic or fermionic harmonic oscillators are employed. This is always possible for representations whose Young diagram is of one-row or one-column form. We remark that these types of representations are relevant for the integrable spin chain that appears in the spectral problem of planar $\mathcal{N}=4$ Super Yang-Mills. One again finds that the solution to the YBE is expressed as a gamma function depending on central elements. We will see in the following that gamma functions also provide a solution for more complicated cases.

After these initial considerations, let us next proceed to the generic case. Here $\bar{I}$ is any set with $q\equiv|\bar{I}|$.  We again look for a solution of \eqref{eqIlookat} depending only on the Casimir operators of $\mathfrak{gl}(\bar{I})$. Then both sides of \eqref{eqIlookat} can be seen as elements of the space $\mathcal{X}_{\bar{I}}$ spanned by 
\begin{equation}
(J^k)^{\dot a}_{b}=J^{\dot{c}_{1}}_{b}J^{\dot{c}_2}_{\dot{c}_1}\ldots J^{\dot a}_{\dot{c}_{k-1}}\,,\qquad\qquad k=1,2,\ldots\, ,
\end{equation}
with coefficients in the center of the universal enveloping algebra $\mathcal{U}(\alg{gl}(\bar{I}))$. An important observation is that these coefficients have very non-trivial commutation relations with elements of $\mathcal{X}_{\bar{I}}$. It is therefore not straightforward to solve \eqref{eqIlookat}. In the above warm-up exercise the key simplifying feature was that the space $\mathcal{X}_{\bar{I}}$ is one-dimensional. In the generic case, thanks to the non-commutative Cayley-Hamilton theorem reviewed in Appendix  \ref{app:Reving}, the space $\mathcal{X}_{\bar{I}}$ has dimension $q$ -- it is finite dimensional. This may be easily seen by multiplying the relation \eqref{noncommCHbis} (with dotted indices in place of the capital ones) from the left by $J^{\dot b}_{a}$
\beq
\label{noncommCHtris}
\left(J^{\,q+1}\right)^{\dot{a}}_{a}=\,\sum_{k=1}^{q}\,\left(J^{\,k}\right)^{\dot{a}}_{a}\,a_k(\hat \ell_1,\dots,\hat \ell_q)\,.
\eeq
The quantities $\hat\ell_1,\ldots,\hat\ell_q$ play an important role in the following. The reader may find more details in Appendix \ref{app:Reving} and Section \ref{sec:analytic}.
In analogy to \eqref{noncommCHtris} one can reduce any higher power of $J$ and express all of them in the basis given by 
\beq
\label{badbasis}
\left(J^{\,k}\right)^{\dot{a}}_{a} \,,\qquad  \qquad \,\,\,k=1,\dots,q\,.
\eeq
However, it turns out that this is not a very convenient basis. We are rather looking for a basis
\beq
\label{goodbasis}
\left(X^{k}\right)^{\dot{a}}_{a}\,,\qquad \qquad\,\,\, k=1,\dots,q\,,
\eeq	
such that for any element $\alg{a}^L$ in the center of $\mathcal{U}(\alg{gl}(\bar{I}))$ there exists another element $\alg{a}^R$ in the center of $\mathcal{U}(\alg{gl}(\bar{I}))$ for which
\beq
\label{separeted}
\alg{a}^L\,\left(X^{k}\right)^{\dot{a}}_{a}=\left(X^{k}\right)^{\dot{a}}_{a}\,\alg{a}^R\,.
\eeq
Finding such a basis is equivalent to the diagonalization problem of a family of commuting $q\times q$ matrices, which simplifies to the diagonalization problem of only one such matrix. The simplest non-trivial element of the center of $\mathcal{U}(\alg{gl}(\bar{I}))$ to be considered is the quadratic Casimir operator $\hat{C}_2=J^{\dot c}_{\dot d}J^{\dot d}_{\dot c}$. One easily derives its action on the basis \eqref{badbasis}
\beq
\label{prec2}
\hat{C}_2\,\left(J^{\,k}\right)^{\dot{a}}_{a}=2\,\left(J^{\,k+1}\right)^{\dot{a}}_{a}\,+\left(J^{\,k}\right)^{\dot{a}}_{a}\,(\hat{C}_2\,+\,q)\,.
\eeq
Thanks to \eqref{noncommCHtris}, the determination of the basis \eqref{goodbasis} reduces to the diagonalization of the single matrix $2\,\mathbf{M}+(C_2+q)\,\mathbf{I} $, where
\begin{equation}\label{O.matrix}
\mathbf{M}\equiv\left(
 \begin{tabular}{ccccc}
 $0$&0&\ldots &0 & $a_{1}$\\
 1&$0$&\ldots&0&$a_{2}$\\
 $\vdots$&$\vdots$&$\ddots$&$\vdots$&$\vdots$\\
 0&0&\ldots&$0$&$a_{q-1}$\\
 0&0&\ldots&1&$a_q$
 \end{tabular}
 \right).
\end{equation}
The elements $a_i$ in the last column of $\bf M$ are symmetric functions of the $\hat \ell_i$ appearing in \eqref{noncommCHtris}. Their explicit form is given in \eqref{aofell}. 
We immediately conclude that after diagonalization one has
\beq
\label{C2exchange}
\hat{C}_2\,\,\left(X^{k}\right)^{\dot{a}}_{a}=\left(X^{k}\right)^{\dot{a}}_{a}\,\left(\hat{C}_2+q+2\,\hat{\ell}_k \right),
\eeq
where the $\hat{\ell}_k$ are elements of the center of $\mathcal{U}(\alg{gl}(\bar{I}))$ (see formula \eqref{elli}). We stress that up to an overall normalization the basis $X^{k}$ is uniquely fixed once $\bf M$ is diagonalized. This means that the basis elements $X^{k}$ have good exchange properties with \emph{all} higher Casimir operators, and therefore with any element of the center of $\mathcal{U}(\alg{gl}(\bar{I}))$, and thus all $\hat{\ell}_i$. The explicit form of the generators in \eqref{badbasis} in terms of the basis \eqref{goodbasis} is given by \eqref{JofX}. 

Let us elaborate on the role of the $\hat \ell_i$ and their relations to the $\mathfrak{gl}(\bar{I})$ Casimir operators \begin{equation}
\hat{C}_i=J_{a_i}^{a_1}J_{a_1}^{a_2}\ldots J_{a_{i-1}}^{a_i}\,.
\end{equation}
The operators $\hat \ell_i$ act by scalar multiplication on the elements of any $\mathfrak{gl}(\bar{I})$ irreducible representation. For highest weight representations their action may be determined by acting with the $\mathfrak{gl}(\bar{I})$ generators on $\mathfrak{gl}(\bar{I})$ highest weight states 
\beq\label{hwscond}
J^{\dot{c}}_{\dot{d}}\,|h.w.s.\rangle\,=0,\qquad \text{for}\,\,\,\,\dot{c}<\dot{d} \,,\qquad\mbox{and}\,\qquad \,J^{\dot{c}}_{\dot{c}}\,\,|h.w.s.\rangle\,=\lambda_{\dot{c}}\,|h.w.s.\rangle\,.
\eeq
In this case the operators $\hat{\ell}_i$ act as {\it shifted weights}, and we deduce\footnote{For convenience we rewrite $\lambda_{\dot{c}_i}$ as $\lambda_{i}$. This rewriting manifests the fact that the natural ordering in the set $\{1,2\dots,n\}$ induces an ordering on the set $\bar{I}$. The same ordering is used in \eqref{hwscond}.} $\hat{\ell}_i|h.w.s.\rangle=(\lambda_i-i+1)|h.w.s.\rangle$. On a given irreducible representation of $\mathfrak{gl}(n)$ the $\mathfrak{gl}(\bar{I})$ shifted weights $\hat \ell_i$ should be understood as {\it operatorial shifted weights}.
 Upon acting on a highest weight state as in \eqref{hwscond} any polynomial in the $\mathfrak{gl}(\bar{I})$ Casimir operators can be rewritten as polynomial in the shifted weights $\hat \ell_i$. This procedure provides an isomorphism between the center of $\alg{gl}(\bar{I})$ and the universal enveloping algebra of the Cartan subalgebra of $\alg{gl}(\bar{I})$. This is known as the Harish-Chandra isomorphism, see for example \cite{molevbook} and references therein. In particular one gets (see \cite{Zhelobenko})
\begin{equation}
\label{Zhelobenkoeq}
\hat{C}_i=\sum_{k=1}^{q}\prod_{j\neq k}\left(1+\frac{1}{\hat{\ell}_k-\hat{\ell}_j}\right)\,(\hat{\ell}_k)^i\,.
\end{equation}
It allows us to write down the commutation relations between $X^k$ and $\hat{\ell}_i$ operators as
\beq
\label{toproove}
\hat{\ell}_i\,\,\left(X^{k}\right)^{\dot{c}}_{c}=\left(X^{k}\right)^{\dot{c}}_{c}\,\left(\delta_{ik}\,\,+\,\,\hat{\ell}_i \right),
\eeq
which is consistent with \eqref{C2exchange}. The exchange relations \eqref{toproove} are entirely fixed considering the analog of \eqref{C2exchange} for higher Casimir operators together with \eqref{Zhelobenkoeq}. 

We are now ready to solve  \eqref{eqIlookat}. We use the relation \eqref{toproove} and the relation between $J^k$ and $X^k$, which we established in the process of diagonalization of the matrix $\bf M$. This leads to the set of equations  
\begin{equation}
\mathcal{R}_{0,I}(z)\, (z-\sfrac{q}{2}-\hat{\ell}^{\bar I}_k+1)\, \left(X^{k}\right)^{\dot{c}}_{c} =\left(X^{k}\right)^{\dot{c}}_{c}\, \mathcal{R}_{0,I}(z)\,,
\end{equation}
which are direct analogs of the $|\bar{I}|=1$ case \eqref{Ibaris1}. We added an extra label $\bar I$ to $\hat{\ell}_k$ in order to emphasize that these are shifted weights of the subalgebra $\mathfrak{gl}(\bar I)$. The solution for $\mathcal{R}_{0,I}(z)$ is then given by
\begin{equation}
\label{R0good!}
\mathcal{R}_{0,I}(z)=\,\rho_I(z)\, \prod_{k=1}^q\,\Gamma(z-\sfrac{q}{2}-\hat{\ell}^{\bar I}_k+1)\,.
\end{equation}
The quantity $\rho(z)$ is an arbitrary function of the spectral parameter. Moreover, it may be an arbitrary periodic function of $\hat{\ell}_k$ of period one. We stress that  \eqref{R0good!} is a symmetric function
in the variables $\hat{\ell}^{\bar I}_{1}\,,\dots, \hat{\ell}^{\bar I}_q$. This is no coincidence since the Weyl group of $\alg{gl}(\bar{I})$ acts on these variables by permutations. The expression for $\mathcal{R}_{0,I}(z)$ is universal and independent of the existence of highest (or lowest) weight states.

To summarize, \eqref{R0good!} together with the factorized expression \eqref{R0tobe} yields the solution to the YBE \eqref{YBE.expl}. We thus found the building blocks for the Q-operators for the spin chain with any $\mathfrak{gl}(n)$ representation $\Lambda$ in quantum space. Before proceeding to the explicit construction of the corresponding novel Q-operators we will first prove some very important fusion relations.

%%%%%%%%%%%%%%%%%%%%%%%%%%%%%%%%%%%%%%%%%%%%%%%%%%%%%%%%%%%%%%%%%
\subsection{Fusion of Lax Operators}
\label{sec:fusionrev}
%%%%%%%%%%%%%%%%%%%%%%%%%%%%%%%%%%%%%%%%%%%%%%%%%%%%%%%%%%%%%%%%%

In this section we present the factorization formula of the Lax operators that leads to the full hierarchy of functional relations. This procedure was exhaustively discussed in \cite{Bazhanov:2010jq} for the fundamental representation of $\alg{gl}(n)$ in quantum space. It provides a solid starting point for the factorization, based on a precise relation between certain degenerate representations of the Yangian algebra. In particular, in the present language, the structure of the auxiliary space is independent of the one of quantum space. Therefore the same functional relations remain valid for arbitrary representations $\Lambda$. Many more details and results on this issue are provided in \cite{thesiscarlo}.

In generalization of \cite{Bazhanov:2010jq} we obtain the fusion relation for non-intersecting sets $I\cap J=\varnothing$ as
\begin{equation}
\label{factJfull}
\Rli^{[1]}(z+\lambda+\sfrac{p_2}{2})\,\mathcal{R}_{ J}^{[2]}(z-\sfrac{p_1}{2})=\mathcal{S}\,{\mathcal{R}^{[1']}_{ {I \cup J}}}(z)\,\mathcal{R}^{[2']}_{\Gbb} \, \mathcal{S}^{-1},
\end{equation}
with 
\begin{equation}
\mathcal{R}^{[2']}_{\Gbb}\,=\,\exp\,\left(-\,\bar{\mathcal{J}}^{a}_{\dot{a}}\,\left(\osca^{[2']}\right)_{a}^{\dot{a}}\right)\,,
\end{equation}
and $\mathcal{S}$ given by formulas (3.8)-(3.10) in \cite{Bazhanov:2010jq}.
In this section dotted indices refer to the subset $J$, and we define $p_1=|I|$, $p_2=|J|$ and $p=p_1+p_2$. The similarity transformation $\mathcal{S}$ acts purely in auxiliary space and disentangles the oscillators, making manifest that ${\mathcal{R}_{ {I \cup J}}}(z)$ and $\mathcal{R}_{\Gbb}$ act in different auxiliary spaces denoted by $[1']$ and $[2']$, respectively.
After a careful analysis of \eqref{factJfull} one concludes that  $\mathcal{R}_{{I \cup J}}(z)$ 
is of the form \eqref{R0tobe} with
\begin{equation}
\label{R0fused}
\mathcal{R}^{[1']}_{0\,,I\cup J}=e^{(\bar{\osca}^{[1']})^{\dot{c}}_c\,J_{\dot{c}}^c}\,\mathcal{R}_{0\,,I}(z_1)\,e^{-(\osca^{[1']})_{\dot{c}}^c\,J^{\dot{c}}_c}\,\mathcal{R}_{0\,,J}(z_2)\,e^{-(\bar{\osca}^{[1']})^{\dot{c}}_c\,J_{\dot{c}}^c}\,.
\end{equation}
Via tedious but straightforward calculations one sees that $\mathcal{R}_{0\,,I\cup J}(z)$ satisfies \eqref{eqIlookat} with $\bar{\mathcal{J}}^{a}_{b}$ in the Holstein-Primakoff form built from oscillators $\osca^{[1']}$ and $\bar\osca^{[1']}$ (see formula (3.12) in  \cite{Bazhanov:2010jq} for their detailed form). It means that $\mathcal{R}_{I\cup J}(z)$ is a solution of the YBE \eqref{YBE.expl} with the $\mathfrak{gl}(I\cup J)$ representation in $\mathcal{A}_{I\cup J}$ given by weights $\tilde\Lambda=(\underbrace{\lambda}_{p_1},\underbrace{0}_{p_2})$.

Following the same line of reasoning as in \cite{Bazhanov:2010jq}, we can summarize this section with the  factorization formula for Lax operators
\begin{equation}\label{factorization}
\kappa_{I}(z)\mathcal{R}_{\{i_1\}}(z+\lambda_{1}')\,\mathcal{R}_{\{i_2\}}(z+\lambda_{2}')\,\ldots\, \mathcal{R}_{\{i_p\}}(z+\lambda_{p}')=\mathcal{S}_I \,\mathcal{R}^{+}_{ I}(z\,|\,\Lambda_p)\,\mathcal{R}_{\mathbb{G}_I}\,\mathcal{S}_{I}^{-1}\,,
\end{equation}
where $I=\{i_1,\ldots,i_p\}$ and $\lambda_{j}'=\lambda_{j}+\sfrac{p-2j+1}{2}$, and $\kappa_I$ depends on the normalization of $\mathcal{R}_{I}(z)$. Here we introduced an infinite dimensional highest weight representation $\pi^{+}_{\Lambda_p}$ of the algebra $\mathfrak{gl}(p)$ with weights $\Lambda_p=(\lambda_j)_{j=1,\ldots,p}$ and
\begin{equation}
\mathcal{R}^{+}_{ I}(z\,|\,\Lambda_p)=\pi^{+}_{\Lambda_p}\left[ \mathcal{R}_{ I}(z)\right].
\end{equation}
The form of \eqref{factorization} is exactly the same as the one obtained for the fundamental representation in quantum space. As in the latter case, when supplemented by the known relation between finite-dimensional and infinite-dimensional modules (the so-called Bernstein-Gelfand-Gelfand (BGG) resolution) it suffices to derive the full hierarchy of  functional relations. This means that the functional relations will be exactly the same as in the earlier case. The difference appears merely in the analytic properties of the Q-operators, which are a direct consequence of formula \eqref{R0good!}. In the following section we shall take a closer look at the analytic properties of $\Rli(z)$.

%%%%%%%%%%%%%%%%%%%%%%%%%%%%%%%%%%%%%%%%%%%%%%%%%%%%%%%%%%%%%%%%%%%%%%%
\subsection{Analytic Properties of Lax Operators}
\label{sec:analytic}
%%%%%%%%%%%%%%%%%%%%%%%%%%%%%%%%%%%%%%%%%%%%%%%%%%%%%%%%%%%%%

Before we proceed with the construction of Q-operators for general lowest/highest weight $\alg{gl}(n)$ spin chains we would like to investigate the analytic structure of the new solutions to the YBE as found in Section \ref{sec:oscpart}. This analytic structure will be reflected in the form of the Q-operators, which will be crucial in Section \ref{sec:exappl}. We want to stress that it follows directly from our operatorial construction of Q-operators. Three types of $\mathfrak{gl}(n)$ representations $\Lambda$ will be distinguished in quantum space: compact representations, non-compact representations with a lowest/highest weight state, and non-compact representations without a lowest/highest weight state. The difference between these three cases is reflected in the structure of the eigenvalues of the $\hat\ell$ operators entering \eqref{R0good!}. Their precise spectrum crucially depends on the representation under study and can be extracted in a case-by-case analysis. However, a common feature of all representations is that the eigenvalues of $\hat\ell^{\bar I}$ for a fixed set $I$ are separated by integers.   

For the compact case the spectra of all $\hat\ell$ are bounded from both above and below. This means that there exists a normalization such that all of the $\Rli(z)$ have polynomial matrix elements. This information is essential when deriving Bethe equations. For non-compact representations with a lowest/highest weight state some of the $\hat\ell^{\bar{I}}$ for a fixed set $I$ are not bounded from above. 
Because of \eqref{R0good!}, it is not possible to find a normalization for which the $\Rli(z)$ is polynomial -- its matrix elements are meromorphic functions. Let us elaborate on this point with the simple example of the spin $-\sfrac{1}{2}$ representation of $\mathfrak{gl}(2)$, which is a unitary representation of $\mathfrak{u}(1,1)$. We may realize it with the use of two oscillators, and define a basis in representation space by
\begin{equation}\label{minus12basis}
| M \rangle =(\bar \osca \bar \oscb)^M |0\rangle\,,\qquad M=0,1,\ldots\, .
\end{equation}
The generators of $\mathfrak{gl}(2)$ can be written with use of oscillators $\bar\osca,\bar\oscb$ as\footnote{Algebraically, this could also be rewritten by a ``particle-hole transformation'' in the standard form $J_{a}^{b}=\bar{\osca}^{b}\osca_a$. To get the appropriate real form $\mathfrak{u}(1,1)$, however, we should use the above form, and impose the reality conditions $\bar \osca=\osca^{\dagger}$ and $\bar \oscb=\oscb^{\dagger}$.}
\begin{eqnarray}
&&J_{1}^{1}=\bar{\osca} \osca\,,\qquad J_{1}^{2}=-\osca\oscb\,,\\
&&J_{2}^{1}=\bar{\osca}\bar{\oscb}\,,\qquad J_{2}^{2}=-1-\bar{\oscb}\oscb\,.
\end{eqnarray}
We have to consider two cases: $I=\{ 1 \}$ or $I=\{ 2\}$. In both of them we are interested in $\mathfrak{gl}(1)$ subalgebras of $\mathfrak{gl}(2)$. We notice that $\hat{\ell}^{\{ 1\}} |M\rangle=J_{1}^{1}|M\rangle= M|M\rangle$  is not bounded from above while $\hat{\ell}^{\{ 2\}} |M\rangle=J_{2}^{2}|M\rangle=-(1+ M)|M\rangle$ is. This means that we can find a normalization for which $\mathcal{R}_{\{ 1\}}(z)$, depending on $\hat{\ell}^{\{2\}}$, is polynomial. This is not possible, however, for $\mathcal{R}_{\{ 2\}}(z)$, which is a function of $\hat{\ell}^{\{ 1\}}$.

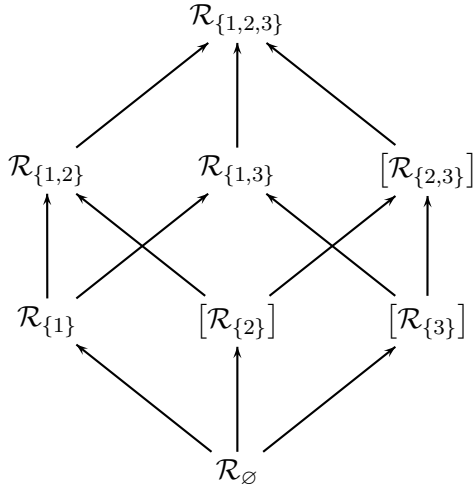
\begin{figure}[t]
\begin{center}
\begin{pspicture}(8,6)
\rput(4,0){\rnode{A0}{$\mathcal{R}_{\varnothing}$}}
\rput(1.5,2){\rnode{A1}{$\mathcal{R}_{\{1\}}$}}
\rput(4,2){\rnode{A2}{$\left[\mathcal{R}_{\{2\}}\right]$}}
\rput(6.5,2){\rnode{A3}{$\left[\mathcal{R}_{\{3\}}\right]$}}
\rput(1.5,4){\rnode{A12}{$\mathcal{R}_{\{1,2\}}$}}
\rput(4,4){\rnode{A13}{$\mathcal{R}_{\{1,3\}}$}}
\rput(6.5,4){\rnode{A23}{$\left[\mathcal{R}_{\{2,3\}}\right]$}}
\rput(4,6){\rnode{A123}{$\mathcal{R}_{\{ 1,2,3\}}$}}
\ncline[ArrowInside=->,ArrowInsidePos=1.0,nodesep=0.1]{A0}{A1}
\ncline[ArrowInside=->,ArrowInsidePos=1.0,nodesep=0.1]{A0}{A2}
\ncline[ArrowInside=->,ArrowInsidePos=1.0,nodesep=0.1]{A0}{A3}
\ncline[ArrowInside=->,ArrowInsidePos=1.0,nodesep=0.1]{A1}{A12}
\ncline[ArrowInside=->,ArrowInsidePos=1.0,nodesep=0.1]{A1}{A13}
\ncline[ArrowInside=->,ArrowInsidePos=1.0,nodesep=0.1]{A2}{A12}
\ncline[ArrowInside=->,ArrowInsidePos=1.0,nodesep=0.1]{A2}{A23}
\ncline[ArrowInside=->,ArrowInsidePos=1.0,nodesep=0.1]{A3}{A13}
\ncline[ArrowInside=->,ArrowInsidePos=1.0,nodesep=0.1]{A3}{A23}
\ncline[ArrowInside=->,ArrowInsidePos=1.0,nodesep=0.1]{A12}{A123}
\ncline[ArrowInside=->,ArrowInsidePos=1.0,nodesep=0.1]{A13}{A123}
\ncline[ArrowInside=->,ArrowInsidePos=1.0,nodesep=0.1]{A23}{A123}
\end{pspicture}
\caption{Hasse diagram for the unitary highest weight representations of $\mathfrak{u}(1,2)$ algebra. The brackets $[~~]$ indicate that the enclosed Lax operator has non-polynomial matrix elements. If the bracket is absent, one may normalize the Lax operator such that its matrix elements are polynomial.}
\label{Hasse12}
\end{center}
\end{figure}

Considering the appropriate reality conditions, a similar analysis can be applied to oscillator representations of $\mathfrak{u}(n,m)$, or even more generally to any unitary highest weight representation of $\mathfrak{u}(n,m)$. The case of $\mathfrak{u}(1,2)$ is given in Figure \ref{Hasse12}, where all $\mathcal{R}$-matrices are depicted on the Hasse diagram. Similar patterns can be obtained for a generic highest weight infinite-dimensional representation of $\mathfrak{gl}(n)$.
For representations without the lowest/highest weight state none of the $\hat\ell$ operators is bounded from above or below and therefore none of R-matrices can be polynomial.

%%%%%%%%%%%%%%%%%%%%%%%%%%%%%%%%%%%%%%%%%%%%%%%%%%%%%%%%%%%%%%%%%
\subsection{Construction of Q-operators}
\label{sec:Qconstruction}
%%%%%%%%%%%%%%%%%%%%%%%%%%%%%%%%%%%%%%%%%%%%%%%%%%%%%%%%%%%%%%%%%%%%%%%%
Let us now proceed with the construction of the Q-operators for arbitrary representations of $\alg{gl}(n)$. We closely follow \cite{Bazhanov:2010jq}, where the trace-construction was studied for the fundamental representation of $\alg{gl}(n)$ in quantum space. There we had introduced more general chain operators $\textbf X_{I}(z)$,  whose building blocks were the Lax matrices in \eqref{Rfi}. The notion of X-operators can be easily generalized to the case of an arbitrary representation in quantum space by the simple substitution $\mathbf{L}_I \to \Rli$. To keep formulas short, we exclusively focus in the following on the case of Q-operators.

The quantum space $\mathcal{V}$ of the $\alg{gl}(n)$ spin chain with the local state space $V$ being in the representation $\Lambda$ is an $L$-fold tensor product space, cf.~\eqref{quantumspace}.
To construct the Q-operators we consider the monodromy matrices 
\begin{equation}\label{monodromy}
 \textbf M_I(z)=\mathcal D_I\,\Rli(z)\otimes \Rli(z)\otimes\ldots\otimes \Rli(z)
\end{equation} 
acting on 
\begin{equation}
\mathcal{V}=\left(V^{\otimes L} \right)\otimes \mathcal A_I\,.
\label{monodromyspace}
\end{equation}
The local building blocks of the $ \textbf M_I$ are the R-matrices $\Rli(z)$ given in (\ref{R0tobe}), which were obtained from the YBE (\ref{YBE.expl}). While the various R-matrices in (\ref{monodromy}) act on distinct spaces $V$ they all share the same auxiliary space $\mathcal A_I$. The boundary twist operator $\mathcal D_I$ acts only in the auxiliary space. Its explicit form is given by
\begin{equation}\label{D-exp}
\mathcal D_I=\exp\Big\{-
i \,\Phi^{\dot b}_a \,\osch_{\dot b}^{a} \Big\}\,,
\end{equation}
where $\Phi^{\dot b}_a=\Phi_a-\Phi_{\dot b}$ and $\osch_{\dot b}^{a}=\bar{\osca}^{\dot b}_{ a} \osca_{\dot b}^{ a}+\sfrac{1}{2}$ (no summation).
The Q-operators are defined by taking the traces of the monodromy matrices (\ref{monodromy}) as
\begin{equation}\label{Q-def}
\textbf{Q}_I(z)=e^{i z\,(\,\sum_{a\in I}\Phi_a)}\,\widehat{\text{Tr}}_{{\mathcal H}^{(I,\bar I)}} 
\big\{\/{\bf M}_I(z)\big\}\,,
\end{equation}
where, following the notation of \cite{Bazhanov:2010jq}, the normalized trace is defined by 
\begin{equation}
 \widehat{\text{Tr}}_{\mathcal H}\big\{e^{-i\Phi\textbf h}\star\big\}\,=\frac{\text{Tr}_{\mathcal H}
\big\{e^{-i\Phi\textbf h}\star\big\}}{\text{Tr}_{\mathcal H}\big\{e^{-i\Phi\textbf h}\big\}}\,.
\end{equation} 
An important feature of the normalized trace is that it is completely determined by the commutation relations \eqref{oscillators} and the cyclic property of the trace.
Recall that, thanks to the underlying YBE, all Q-operators above form, for a fixed quantum space, a commuting family of operators.
They belong to a larger family of commuting operators $\Xbf(z)$, which includes the standard transfer matrices as special cases.
The next task is to diagonalize the Q-operators. This is done using the QQ-relations, to be derived in the next section, in conjunction with the analytic structure of the eigenvalues of the Q-operators.

%%%%%%%%%%%%%%%%%%%%%%%%%%%%%%%%%%%%%%%%%%%%%%%%%%%%%%%%%%
\subsection{Functional Relations and Bethe Equations}
\label{sec:exappl}
%%%%%%%%%%%%%%%%%%%%%%%%%%%%%%%%%%%%%%%%%%%%%%%%%%%%%%%%%%%%%%%%%%%%%%%%%

Applying the reasoning from \cite{Bazhanov:2010jq} to the fusion relations of Section \ref{sec:fusionrev} one obtains two important relations between Q-operators, allowing to diagonalize the problem with the use of Bethe equations. All Q-operators can be expressed using only partonic Q-operators. These correspond to the sets $I=\{ i\}$, containing only a single index $i=1,\ldots, n$. The formula is given by the following determinant\footnote{One can check that this holds if we take $\rho_{I}(z)=1$ in \eqref{R0good!} for all $I$. Another interesting normalization is 
\begin{equation}
\rho_{I}(z)=\prod_{i=q+1}^{n}\Gamma(z+i-q/2)\left(\prod_{i=1}^{n}\Gamma(z-\ell_i-q/2+1)\right)^{-1}\,,
\end{equation} 
in which case all Q-operators are polynomial for compact representations.}
\begin{equation}
\Delta_I(\Phi) \Qop_I(z)=\det || \Qop_{a_i}(z-j+\sfrac{p+1}{2})||_{1\leq i,j\leq |I|}\,,
\end{equation}
where $\Delta_I$ depends only on the twist angles defined in Section \ref{sec:Qconstruction}. It reads
\begin{equation}
\Delta_I(\Phi)=\prod_{1\leq i,j\leq |I|} 2i\sin\left(\sfrac{\Phi_{a_i}-\Phi_{a_j}}{2}\right).
\end{equation}
As a direct consequence of these relations one can write down the QQ-relations, see also \cite{Tsuboi:2009ud}, as
\beq
\label{QQgene}
\Delta_{\{a,b\}}(\Phi)\,\Qop_{I\,\cup\,a\,\cup\,b}(z)\,\Qop_{I}(z)=\,\Qop_{I\,\cup\,a}(z-\half)\,\Qop_{I\,\cup\,b}(z+\half)-\,\Qop_{I\,\cup\,b}(z-\half)\,\Qop_{I\,\cup\,a}(z+\half)\,.
\eeq
Just as in the earlier case of the fundamental representation in quantum space we can associate all Q-operators to the vertices of a Hasse diagram, which for $\alg{gl}(n)$ takes the form of an $n$-dimensional hypercube. Each of its faces corresponds to precisely one relation \eqref{QQgene}. With every path along the edges of the diagram leading from $\Qop_{\varnothing}(z)$ to $\Qop_{\{1,\ldots,n\}}(z)$ we can associate a system of Bethe equations. The different paths correspond to the different sequences of ascending sets $\varnothing=I_0\subset I_1\dots \subset I_n= \{1,2\dots,n\}$. Clearly there are $n!$ such paths. The functional form of Bethe equations is then given by
\begin{equation}\label{formalBETHE3}
-1=\frac{{\rm Q}_{I _{i-1}}(\specbaz_k^{I_i}-\half)}{{\rm Q}_{I_{i-1} }(\specbaz_k^{I _i}+\half)}\,
\frac{{\rm Q}_{I_i}(\specbaz_k^{I_i}+1)}{{\rm Q}_{I_i}(\specbaz_k^{I_i}-1)}\,
\frac{{\rm Q}_{I_{i+1}}(\specbaz_k^{I_i}-\half)}{{\rm Q}_{I_{i+1}}(\specbaz_k^{I_i}+\half)}\,,
\end{equation}
where $z_{k}^{I_i}$ are the roots of ${\rm Q}_{I_i}(z)$, and we denoted the eigenvalues of the operator $\Qop_{I}(z)$ by ${\rm Q}_{I}(z)$.

So far we did not use any information about the analytic structure of the Q-operators, apart from the fact that there exists at least one root of ${\rm Q}_I(z)$. It is needed in order to rewrite \eqref{formalBETHE3} into the standard explicit form of Bethe equations. This is possible whenever all operators $\Qop_{I_i}(z)$ for $i=0,\ldots,n$ on the chosen path can be normalized to be polynomials (see the discussion in the Section \ref{sec:analytic}). Let us start from compact representations in quantum space. Then for all $\Qop_I(z)$ we can write their eigenvalues in the form
\beq
\label{anal}
{\rm Q}_I(z)\,=\,e^{\,z\,\left(\sum_{i\in\,I}\,\Phi_i\right)}\,\,\left( \prod_{a=1}^q\,\Gamma(z-\sfrac{q}{2}-\ell_a+1)\right)^L\,\prod_{k=1}^{m_I}\,\left(z\,-\,z^{I}_k \right)\,.
\eeq 
Substituting this form into \eqref{formalBETHE3} we end up with
\begin{eqnarray}\label{betheequations}
e^{i(\Phi_{a_{i+1}}-\Phi_{a_i})}\left( \frac{z^{I_i}_{l}-\sfrac{n-i-1}{2}-\ell_{a_{n-i+1}}-\sfrac{1}{2}}{z^{I_i}_{l}-\sfrac{n-i-1}{2}-\ell_{a_{n-i}}+\sfrac{1}{2}}\right)^L=&&\\\nonumber
&& \hspace{-4cm}=\prod_k\frac{z_l^{I_i}-z_k^{I_{i-1}}-\sfrac{1}{2}}{z_l^{I_i}-z_k^{I_{i-1}}+\sfrac{1}{2}}\prod_{k\neq l}\frac{z_l^{I_i}-z_k^{I_i}+1}{z_l^{I_i}-z_k^{I_i}-1}\prod_k\frac{z_l^{I_i}-z^{I_{i+1}}-\sfrac{1}{2}}{z_k^{I_i}-z^{I_{i+1}}+\sfrac{1}{2}}\,.
\end{eqnarray}
These equations appeared for the first time in \cite{Kulish:1983rd}, where they were derived by the algebraic Bethe ansatz technique. For compact representations, due to Weyl permutation symmetry, we have that for fixed $|I|$ all $\Qop_I(z)$ are isospectral. This means that all $n!$ systems of Bethe equations have exactly the same form. This is not true for non-compact representations -- Weyl symmetry is broken, a fact which leads to different analytic properties of the Q-operators. For representations with a lowest/highest weigh state, however, there still exist paths on the Hasse diagram for which all $\Qop_I(z)$ connected by the path can be normalized to be polynomial. In that case the formula \eqref{anal} holds true and we are again able to rewrite \eqref{formalBETHE3} into the standard form \eqref{betheequations}. This means that, using the Q-operator approach, for lowest/highest weight compact or non-compact representations it is always possible to write polynomial Bethe equations for spin chains.

%%%%%%%%%%%%%%%%%%%%%%%%%%%%%%%%%%%%%%%%%%%%%%%%%%%%%%%%%%%%%%%%%%%
\subsection{Q-operators in use}
\label{sec:qopinuse}
%%%%%%%%%%%%%%%%%%%%%%%%%%%%%%%%%%%%%%%%%%%%%%%%%%%%%%%%%%%%%%%%%%%
In the previous sections we obtained new solutions of the YBE and used them to construct Q-operators. We furthermore proved operatorial relations between them, and exploited these to derive Bethe equations without the use of a Bethe ansatz. However, for practical purposes such as finding Bethe roots or eigenvectors from Q-operators, it is necessary to find their exact operatorial form by constructing suitable monodromy matrices and explicitly evaluating the trace over their auxiliary space. With increasing length of the spin chain and dimension of the quantum space this becomes a very cumbersome problem. 

For compact representations the Q-operators are $(\mbox{dim}_{\Lambda})^L \times(\mbox{dim}_{\Lambda})^L$ matrices, where $\mbox{dim}_\Lambda$ denotes the dimension of the representation $\Lambda$, and $L$ is the length of the spin chain. We will present some powerful {\it Mathematica}\texttrademark\ code which allows to construct concrete matrix representations of Q-operators for various specific compact representations. It is based on the Gelfand-Tsetlin basis, and serves as a checking device for many properties presented earlier. It may also be used to extract Bethe roots and eigenvectors for given representations. This code can be found in Appendix \ref{app:Qcode}.

For non-compact representations with a lowest/highest weight state the construction of Q-operators is much more involved. The reason is that the representation space is infinite dimensional and therefore the size of the matrices becomes infinite dimensional. It is nevertheless sometimes possible to obtain explicit results, and to evaluate the trace analytically. For example, for the $L=1$ spin chain for any $\mathfrak{gl}(2)$ representation $\Lambda$ in quantum space the Q-operators can be written as
\begin{eqnarray}
\Qop_{\{a\}}(z)=\Gamma(z-J_{\dot a}^{\dot a}+\half) \, _2F_1\left(J^{\dot a}_{\dot a}-\ell_1,J_{\dot a}^{\dot a}-\ell_2;\half-z+J_{\dot a}^{\dot a};\frac{1}{1-e^{i (\phi_a-\phi_{\dot a}) }}\right),
\end{eqnarray}
where $\{a\}\cup\{\dot a\}=\{1,2\}$, and $_2F_1(a,b;c;z)$ stands for the hypergeometric function. One can obtain the same result solving directly the Baxter equation (see e.~g.~\cite{Beccaria:2009rw}). For longer chains it is possible to write Q-operators purely in terms of $\mathfrak{gl}(n)$ generators. However,  we are not aware of any closed formula in that case -- they are written as infinite sums. It is nevertheless possible to extract at least parts of the Q-operators by using the fact that they are block-diagonal matrices with finite-dimensional blocks.

%%%%%%%%%%%%%%%%%%%%%%%%%%%%%%%%%%%%%%%%%%%%%%%%%%%%%%%%%%%%
%%%%%%%%%%%%%%%%%%%%%%%%%%%%%%%%%%%%%%%%%%%%%%%%%%%%%%%%%%%%
\section{Hamiltonians}
\label{sec:hamiltonianchapter}

\subsection{Stating the Problem}
\label{sec:IntroHam}
%%%%%%%%%%%%%%%%%%%%%%%%%%%%%%%%%%%%%%%%%%%%%%%%%%%%%%%%%
The purpose of this section is twofold. Firstly, we provide an explicit and computationally very efficient form of the nearest-neighbor Hamiltonians for those integrable spin chains where the local representation space $V$ in \eqref{quantumspace} is in a so-called {\it generalized rectangular} $\mathfrak{gl}(n)$ representation. It generalizes the famous digamma formula for the $n=2$ case \cite{Babujian:1982ib} as derived from the $\mathfrak{gl}(2)$ R-matrix of Kulish, Reshetikhin and Sklyanin \cite{Kulish:1981gi}, which is widely used in many areas of theoretical physics such as the AdS/CFT correspondence, QCD high-energy scattering problems, and integrable lattice models of condensed matter theory. Secondly, we show that the method we used in Section~\ref{sec:qoperators} for finding the Lax operators for Q-operators is rather universal, and can be used also in other cases. A particularly interesting feature of our result is the appearance of operators corresponding to shifted weights of tensor product representations.

In Section~\ref{sec:qoperators} we used the QISM as a guiding principle, and derived the solution of integrable nearest-neighbor spin chains for arbitrary lowest/highest weight representations of $\mathfrak{gl}(n)$ without employing the Bethe ansatz. This results in (equivalent) sets of Bethe equations, and simultaneously diagonalizes a large class of operators: standard transfer matrices $\Top(z)$, Baxter Q-operators $\Qop(z)$, and mixed forms $\Xop(z)$. In \cite{Tarasov:1983cj} it was shown that integrable nearest-neighbor Hamiltonians belong to the same commuting operator family. As such they are a consequence of the YBE, and thus naturally embedded in the framework of the QISM. To be more precise, in order to extract the Hamiltonian one has to construct the transfer matrix operator $\Top_{\Lambda}(z)$, where the representation of the auxiliary space $\Lambda$ is identical to the one of the local one-site quantum space $V$ of \eqref{quantumspace}. This special transfer matrix contains the information about all conserved {\it local} charges such as the shift operator, the nearest-neighbor Hamiltonian, and a further tower of a total of $L$ higher conserved local charges. In order to extract all of these, one has to expand the transfer matrix in the spectral parameter $z$ around a special regularity point $z=z_*$. One then reads off the various local operators as the coefficients of a Taylor expansion around $z_*$. The regularity point $z_*$ is defined by the requirement that the building block of  $\Top_{\Lambda}(z)$, namely $\mathcal{R}_{\Lambda\Lambda}(z)$, turns into the permutation operator at $z=z_*$ as
\begin{equation}
\mathcal{R}_{\Lambda\Lambda}(z_*)=\mathbf{P}\,.
\end{equation}
In general, the process of finding $\mathcal{R}_{\Lambda\Lambda}(z)$ reduces to solving the appropriate YBE. When the representation $\Lambda$ is fixed it is sufficient to consider the YBE on the tensor product space $\square\otimes \Lambda\otimes\Lambda$, where we denoted the fundamental representation of $\mathfrak{gl}(n)$ by $\square$. For generic representations solving the YBE is a tedious task. However, one may use a shortcut, exploiting the fact that $\mathcal{R}_{\Lambda\Lambda}(z)$ is a $\Lambda\otimes\Lambda$-invariant operator. This means that the following equations are satisfied
\begin{equation}\label{llinv}
[\mathcal{R}_{\Lambda\Lambda}(z), J_{a}^{b}\otimes 1+1\otimes J_{a}^{b}]=0\,,\qquad a,b=1,\ldots, n\,,
\end{equation}
where $J_{a}^{b}$ in both spaces are evaluated in the representation $\Lambda$. For compact representations, for instance, there is a finite and rather small number of independent numerical matrices $\inv_i$ satisfying relation \eqref{llinv}. Then the R-matrix can be written as
\begin{equation}
\mathcal{R}_{\Lambda\Lambda}(z)=\sum \alpha_i(z) \,\inv_i\,,
\end{equation}
where $\alpha_i(z)$ are spectral parameter-dependent functions. Substituting this form of R-matrix into the YBE immediately fixes the functions $\alpha_i(z)$. This way one can find $\mathcal{R}_{\Lambda\Lambda}(z)$ and thus the Hamiltonian density in terms of a series for any fixed representation $\Lambda$ in quantum space. However, the question remains whether there is some closed-form expression for $\mathcal{R}_{\Lambda\Lambda}(z)$ which works for any representation, i.e.~is written purely in terms of the abstract algebra generators. It is known in the literature that the answer is partially affirmative for the case of those representations whose tensor product decomposition is multiplicity free \cite{Gould:2002de}. In that case, using the ``tensor product graph method'', one can write $\mathcal{R}_{\Lambda\Lambda}(z)$ as a sum of projectors, with the coefficients depending on the second Casimir of the representations in the tensor product decomposition of $\Lambda\otimes \Lambda$. However, the form of $\mathcal{R}_{\Lambda\Lambda}(z)$ obtained by using this method is not explicit, as one needs to construct case-by-case the needed projectors. 

The goal of the present Section~\ref{sec:hamiltonianchapter} is then to provide a completely new form of $\mathcal{R}_{\Lambda\Lambda}(z)$ for all rectangular (and therefore multiplicity-free) representations. The new form is both practically efficient as well as suitable for abstract considerations. The resulting R-operator is written in terms of gamma functions of the shifted weights of the tensor product algebra. When expanded in the vicinity of the permutation point $z_*$ it gives a general form of the Hamiltonian density as the sum over digamma functions, in generalization of the $\mathfrak{gl}(2)$ case. In order to find the general form of $\mathcal{R}_{\Lambda\Lambda}(z)$ we proceed as in the previous section, and solve the YBE equation without ever specifying the representation $\Lambda$. The solution is, up to a trivial multiplicative factor, unique, and is in perfect agreement with the one found using the tensor product graph method. 

In order to connect back to Section~\ref{sec:qoperators}, we point out that it is also, in principle,  possible to obtain the transfer matrix $\Top_{\Lambda}(z)$ by means of suitable fusion of Q-operators. This furnishes an alternative in-principle method for deriving the nearest-neighbor Hamiltonian of any spin chain with given representation $\Lambda$ of $V$ in \eqref{quantumspace}. While we do not know how to directly find a ``nice'' formula for $\mathcal{R}_{\Lambda\Lambda}(z)$ from this procedure, fusion nevertheless allows us, using functional relations, to write the dispersion law relating the set of Bethe roots found from the system of Bethe equations to the eigenvalues of the Hamiltonian. The derivation of the dispersion relation along these lines is presented in Section \ref{sec:dispersion}.

%For non-rectangular representations the situation is much more complicated. The reason for that is that the tensor product decomposition is not multiplicity free. Still, one can solve the YBE equation directly, as explained above, and find that there exist two solutions for $\Rll$ for a given representation. It means that there are also two different Hamiltonian densities meaning two distinguished integrable spin chains. We associate this fact with the parity-breaking. An interesting observation is that the spectrum of both spin chains is complex. We did not find any closed formula for the $\Rll$ in the case of non-rectangular representations. It will be easy to see in the following how our derivation of $\Rll$ breaks down in that case.  

%%%%%%%%%%%%%%%%%%%%%%%%%%%%%%%%%%%%%%%%%%%%%%%%%%%%%%%%%
%\subsection{Yang-Baxter Equation for \texorpdfstring{$\Lambda\otimes \Lambda$}{}}
\subsection{Yang-Baxter Equation, Revisited}
\label{sec:Rlam}
%%%%%%%%%%%%%%%%%%%%%%%%%%%%%%%%%%%%%%%%%%%%%%%%%%%%%%%%%
We focus in this section on the $\mathfrak{gl}(n)$ representations $\Lambda$ satisfying the {\it generalized rectangularity condition}
   \begin{equation}\label{JJisJ}
J_{b}^{c}J_{c}^{a}=\alpha J_{b}^{a}+\beta\, \delta_{b}^{a}\,.
\end{equation}
This condition is always satisfied for the $\mathfrak{gl}(2)$ generators as a consequence of the non-commutati-ve Cayley-Hamilton theorem. For $n>2$ the condition \eqref{JJisJ} imposes restrictions on the class of representations considered. Such representations of $\mathfrak{gl}(n)$ shares some important feature with representations of $\mathfrak{gl}(2)$, most notably the tensor product of two such representations as well as their weight diagram are multiplicity free. For finite dimensional representations one can check  that (3.4) is satisfied if and only if the corresponding Young diagram is rectangular. In this case $\alpha=s-a$ and $\beta=s\,a$, where $s$ is the number of columns and $a$ is the number of rows in the Young diagram associated with $\Lambda$. In general one can use the Holstein-Primakoff representation\footnote{To be more precise these representations are exactly of the type entering \eqref{R0fused} in the case in which $\mathcal{R}_{0,I}(z)$ and $\mathcal{R}_{0,J}(z)$ corresponds to minimal representations.} to prove that \eqref{JJisJ} is true in the case
\begin{equation}
\lambda_1=\lambda_2=\ldots=\lambda_a\neq \lambda_{a+1}=\ldots=\lambda_n\, ,
\end{equation} 
where $\lambda_i$ are the Dynkin labels of the representation $\Lambda$ (see also \cite{Green:1971rp}). This is true for the rectangular compact representations where $\lambda_a>\lambda_{a+1}$, and corresponds to non-compact representations in the converse case. 

As was pointed out in Section \ref{sec:IntroHam}, we would like to find the general form of $\mathcal{R}_{\Lambda\Lambda}(z)$ when expressed in terms of the generators of $\mathfrak{gl}(n)$. To this aim we solve the YBE
\begin{equation}\label{YBE.RLL}
\Rfl(z_1)\Rfl(z_2)\Rlam(z_2-z_1)=\Rlam(z_2-z_1)\Rfl(z_2)\Rfl(z_1),
\end{equation}
where $\Rfl(z)$ is given by \eqref{Rfl}. We have also replaced $\mathcal{R}_{\Lambda\Lambda}(z)$ by $\mathfrak{R}(z)$, since we would like to leave the representation $\Lambda$ unspecified for the moment. Assuming  $\mathbf{P}\,\mathfrak{R}(z)\,\mathbf{P}=\mathfrak{R}(z)$, we can rewrite \eqref{YBE.RLL} in the form
\begin{equation}\label{YBE2site}
\left( z(J_{b}^{a}-\tilde{J}_{b}^{a})+(J_{b}^{c}\tilde{J}_{c}^{a}-\tilde{J}_{b}^{c}J_{c}^{a})\right) \Rlam(z)=\Rlam(z)\left( z(J_{b}^{a}-\tilde{J}_{b}^{a})-(J_{b}^{c}\tilde{J}_{c}^{a}-\tilde{J}_{b}^{c}J_{c}^{a})\right),
\end{equation}
where by $J$ and $\tilde{J}$ we denote the generators of the two copies of $\mathfrak{gl}(n)$. 
We now solve this equation using a method very similar to the one in Section \ref{sec:oscpart},  the Cayley-Hamilton theorem again being the main ingredient. This time, however, we use it for the tensor product algebra instead of the subalgebra of $\mathfrak{gl}(n)$. We only sketch here the main steps of the derivation, which goes along the same lines as in Section \ref{sec:oscpart}. Let us introduce a basis by defining
%\footnote{It is a basis of some subspace of the universal enveloping algebra of the tensor product of two $\mathfrak{gl}(n)$ algebras.}
\begin{equation}\label{badbasis2site}
(\mathfrak{J}^{k+1})_{b}^{a}:=(J_{b}^{c_1}-\tilde{J}_{b}^{c_1})(J_{c_1}^{c_2}+\tilde{J}_{c_1}^{c_2})(J_{c_2}^{c_3}+\tilde{J}_{c_2}^{c_3})\ldots(J_{c_{k}}^{a}+\tilde{J}_{c_{k}}^{a}).
\end{equation}
From the Cayley-Hamilton theorem we prove that the space spanned by the $\mathfrak{J}^k$ is finite-dimensio-nal, and that
\begin{equation}
\left(\mathfrak{J}^{\,n+1}\right)^{a}_{b}=\,\sum_{k=1}^{n}\,\left(\mathfrak{J}^{\,k}\right)^{a}_{b}\,\mathfrak{a}_k(A_1,\dots,A_q)\,,
\end{equation}
where $A_i$ are shifted weights of the tensor product representation. After a straightforward calculation one finds the commutation relations of the basis operators \eqref{badbasis2site} with the second Casimir of the tensor product representation
\begin{equation}
\left[C_2,\mathfrak{J}^{k}\right]=4\,\mathfrak{J}^{k+1}-4\,\alpha \,\mathfrak{J}^k\,,
\end{equation}
where $\alpha$ is the constant in the relation \eqref{JJisJ}.
Following the same line of reasoning as before we diagonalize the second Casimir $C_2$ and obtain
\begin{equation}
C_2 \mathfrak{X}_k=\mathfrak{X}_k(C_2+4\hat{A}_k-4\alpha), \qquad k=1,\ldots,n.
\end{equation}
This relation is solved by
\begin{equation}
\left[ \hat{A}_j,\mathfrak{X}_k\right]=\delta_{jk}\mathfrak{X}_k-\delta_{j\bar{k}}\mathfrak{X}_{k}\,, \qquad k=1,\ldots a\,.
\end{equation}
We now have all ingredients necessary to solve \eqref{YBE2site}. Let as rewrite it first in the basis \eqref{badbasis2site}
\begin{equation}
\left((z-\alpha) (\mathfrak{J}^1)_{b}^{a}+(\mathfrak{J}^2)_{b}^{a}\right)\Rlam(z)=\Rlam(z)\left((z+\alpha) (\mathfrak{J}^1)_{b}^{a}-(\mathfrak{J}^2)_{b}^{a}\right).
\end{equation}
After converting to the $\mathfrak{X}_i$ basis and using the fact the constant in \eqref{JJisJ} is given in terms of the tensor product representation weights as
\begin{equation}
\alpha=\frac{\hat{A}_{k}+\hat{A}_{\bar{k}}-1}{2}\,,\qquad \mbox{with } \bar{k}=2a-k+1\,,
\end{equation}
we find
\begin{equation}
\mathfrak{X}_k\left(z +\frac{\hat{A}_k-\hat{A}_{\bar{k}}+1}{2} \right)\Rlam(z)=\Rlam(z)\mathfrak{X}_k\left(z -\frac{\hat{A}_k-\hat{A}_{\bar{k}}+1}{2} \right),\qquad k=1,\ldots,a.
\end{equation}
The solution is then derived to be
\begin{equation}\label{Rllsolution}
\Rlam(z)=\rho(z)\prod_{k=1}^{a}\Gamma\left(z +\frac{\hat{A}_k-\hat{A}_{\bar{k}}+1}{2}\right)\Gamma\left(z -\frac{\hat{A}_k-\hat{A}_{\bar{k}}+1}{2}+1\right).
\end{equation}
The function $\rho(z)$ can be any function of the spectral parameter and any periodic function of the $\hat{A}$. In Appendix \ref{app:TPG} we present the check that this formula indeed agrees with the one found from the tensor product graph method. In the following we will further demand that the R-matrix satisfies the unitarity condition
\begin{equation}\label{unitarity}
\Rlam(z)\Rlam(-z)=1\,.
\end{equation} 
This fixes the normalization of $\Rlam(z)$ to be
\begin{equation}\label{Rllnorm}
\rho(z)=\left(\prod_{k=1}^a \Gamma\left(z +\frac{\hat{A}_k+\hat{A}_{\bar{k}}+1}{2}+2a-k\right)\Gamma\left(z -\frac{\hat{A}_k+\hat{A}_{\bar{k}}+1}{2}-2a+k+1\right)\right)^{-1},
\end{equation}  
unique up to any function satisfying \eqref{unitarity}. In order to apply this result to a specific representation $\Lambda$, we merely need to change the notation back from $\mathfrak{R}(z)$ to $\mathcal{R}_{\Lambda\Lambda}(z)$. Since the expressions \eqref{Rllsolution}, \eqref{Rllnorm} only depend on the shifted weight operators $\hat A_k$ and the spectral parameter $z$, they may be immediately applied to any $\Lambda$.

%\begin{equation}
%\mathcal{R}_{\Lambda\Lambda}(z)=(\pi_{\Lambda}\otimes\pi_{\Lambda})\Rlam(z)\,.
%\end{equation}

%%%%%%%%%%%%%%%%%%%%%%%%%%%%%%%%%%%%%%%%%%%%%%%%%%%%
%\subsection{\texorpdfstring{$\Rll$}{} from \texorpdfstring{$\Rli$}{}}
%%%%%%%%%%%%%%%%%%%%%%%%%%%%%%%%%%%%%%%%%%%%%%%%%%%

%We would like to point out here a curious relation between the solutions of the YBE from which we constructed Q-operators and the solutions we found in the previous section for the rectangular representations in the quantum space. 

%Comparison of decompositions
%\begin{itemize}
%\item Decomposition to the subalgebra
%\begin{eqnarray*}
%\yng(2,2) \to \young(\bullet\bullet,\bullet\bullet)\oplus \young(\bullet\bullet,\bullet\circ)\oplus\young(\bullet\bullet,\circ\circ)\oplus\young(\bullet\circ,\bullet\circ)\oplus\young(\bullet\circ,\circ\circ)\oplus\young(\circ\circ,\circ\circ)
%\end{eqnarray*}
%\item Decomposition of the tensor product
%\begin{eqnarray*}
%\yng(2,2) \otimes \yng(2,2) = \young(\cdot\cdot\bullet\bullet,\cdot\cdot\bullet\bullet)\oplus \young(\cdot\cdot\bullet\bullet,\cdot\cdot\bullet,\circ)\oplus\young(\cdot\cdot\bullet\bullet,\cdot\cdot,\circ\circ)\oplus\young(\cdot\cdot\bullet,\cdot\cdot\bullet,\circ,\circ)\oplus\young(\cdot\cdot\bullet,\cdot\cdot,\circ\circ,\circ)\oplus\young(\cdot\cdot,\cdot\cdot,\circ\circ,\circ\circ)
%\end{eqnarray*}
%end{itemize}

%%%%%%%%%%%%%%%%%%%%%%%%%%%%%%%%%%%%%%%%%%%%%%%%%%%%%%%%%%%%%%%%%%%%
\subsection{Permutation and Hamiltonian Density from the R-matrix}
\label{sec:hamiltonian}
%%%%%%%%%%%%%%%%%%%%%%%%%%%%%%%%%%%%%%%%%%%%%%%%%%%%%%%%%

We next use equation \eqref{Rllsolution} to derive an explicit form of the permutation operator and the Hamiltonian density for the spin chain. The result will again be written in terms of the shifted weights of the tensor product representation. The operators are given, respectively, by the first and the second term of the expansion of $\Rlam(z)$ in the spectral parameter
\begin{equation}
\Rlam(z)=\mathbf{P}(1-z\, \mathcal{H}+\ldots).
\end{equation}
Then the permutation operator is simply $\Rlam(0)$, which from \eqref{Rllsolution} and by using identities for the gamma functions reads
\begin{equation}
\mathbf{P}=\prod_{k=1}^{a}(-1)^{\sfrac{1}{2}(\hat{A}_{k}-\hat{A}_{\bar{k}})}.
\end{equation}
In order to extract $\mathcal{H}$ we take the logarithmic derivative of $\Rlam(z)$ and evaluate it at the permutation point $z=z_*=0$, yielding
\begin{equation}\label{Hdensity}
\mathcal{H}=-\frac{d}{dz}\log \Rlam(z)\Big| _{z=0}\,.
\end{equation}
After straightforward calculations, which are presented in Appendix \ref{app:ham}, we end up with the following form of the Hamiltonian density for the spin chain with the representation in the quantum space satisfying the generalized rectangularity condition \eqref{JJisJ}:
\begin{equation}\label{Hdensity2}
\mathcal{H}=-2 \sum_{k=1}^a\left[ \psi\left(\frac{\hat{A}_k-\hat{A}_{\bar k}+1}{2}\right)- \psi\left(\frac{\hat{A}_{\bar k}+\hat{A}_k+1}{2}+2a-k\right)\right].\\
\end{equation}
It is easy to see that it reproduces the well-known \cite{Babujian:1982ib}, \cite{Faddeev:1996iy} form of the Hamiltonian density for $\mathfrak{su}(2)$ spin chain with spin $s$, namely 
\begin{equation}
\mathcal{H}_{s}=2\psi(s+1)-2\psi\left(\frac{\hat{A}_1-\hat{A}_2+1}{2}\right)=2h(s)-2h(\mathbb{J}).
\end{equation}

Let us pause and make an interesting observation. It is clear from \eqref{Hdensity2} that the value of the Hamiltonian density for a given multiplet on the right hand side of the decomposition depends only on the shape of the Young diagram but not on the algebra at hand -- it is the same for $\mathfrak{gl}(2)$ and for $\mathfrak{gl}(102)$. This means that the energies appearing in the spectrum of the periodic (in particular, all twist angles are zero) spin chain with length $L=2$ are exactly the same for all $\mathfrak{gl}(n)$. Dependence on the rank appears, however, when we calculate the dimensions of multiplets that grow with the rank of the algebra. This fact is not anymore true if we consider longer chains because in that case it is known that the spectrum cannot be fixed completely by symmetry arguments alone.

We recall that, as in \cite{Bazhanov:2010jq}, the nearest-neighbor Hamiltonian commuting with the family of transfer matrices constructed in the previous section is
\begin{equation}
\mathbf{H}=\sum_{i=1}^{L}\mathcal{H}_{i,i+1}\,,\qquad \mbox{with}\qquad \mathcal{H}_{L,L+1}:=e^{i\sum_{a}\phi_a \,J_{a}^{a}(L)}\mathcal{H}_{L,1} e^{-i\sum_{a}\phi_a \,J_{a}^{a}(L)}\,,
\end{equation}
where $\mathcal{H}_{i,i+1}$ is the Hamiltonian density given in \eqref{Hdensity}. The presence of the twist angles $\phi_a$ breaks the $\mathfrak{gl}(n)$ invariance of the chain and allows to probe the ``fine structure'' of the spectrum.
%%%%%%%%%%%%%%%%%%%%%%%%%%%%%%%%%%%%%%%%%%%%%%%%%%%%%%%%%%%%%
\subsection{Energy Formula from Functional Relations}
\label{sec:dispersion}
%%%%%%%%%%%%%%%%%%%%%%%%%%%%%%%%%%%%%%%%%%%%%%%%%%%%%%%%%%%%%
The process of diagonalization of the Hamiltonian for the spin chain using Bethe equations is not complete unless we supplement them with the dispersion relation, which allows to determine the spectrum of the Hamiltonian in terms of the Bethe roots. We show here that for the finite-dimensional representations the operator  $\Top_{\Lambda}(z)$ generating the Hamiltonian  can be expressed solely as a function of a chain of Q-operators connected by a single path in the Hasse diagram. Then, the eigenvalues of the former (energies) can be written in terms of eigenvalues of the latter (functions of the Bethe roots). In order to extract energies from $\Top_\Lambda(z)$ we write for every eigenvalue
\begin{equation}\label{EofT}
E= -\frac{d}{dz}\log T_{\Lambda}(z)|_{z=z_\star}\,.
\end{equation}
We recall that the transfer matrix belongs to a larger family of operators, namely X-operators, and that for the latter we can write, again following \cite{Bazhanov:2010jq},
\begin{equation}\label{XdetQ}
\Delta_{I_k}{{\rm X}}_{I_k}(z_1,\ldots,z_k)=\det_{a,b\in I_k} {{\rm Q}}_a(z_b)\,.
\end{equation}
In \eqref{XdetQ} we omitted the explicit dependence on the spectral parameter, which may be recovered for every ${\rm X}_{I_k}(z)$ by taking the mean value of all the numbers $z$ present in the expression. From the properties of determinants we can write
\begin{equation}\label{Plucker}
\frac{{{\rm X}}_{I_k}(z_0,z_2,\ldots,z_k)}{{{\rm X}}_{I_k}(z_1,z_2,\ldots,z_k)}=\frac{{\rm X}_{I_k}(z_0,z_1,\ldots,z_{k-1})}{{\rm X}_{I_k}(z_1,z_2,\ldots,z_k)}\frac{{\rm X}_{I_{k-1}}(z_2,\ldots,z_k)}{{\rm X}_{I_{k-1}}(z_1,z_2,\ldots,z_{k-1)}}+\frac{{\rm X}_{I_{k-1}}(z_0,z_2,\ldots,z_{k-1})}{{\rm X}_{I_{k-1}}(z_1,z_2,\ldots,z_{k-1})}\,,
\end{equation}
which is is just a version of Pl\"ucker's relations, and thus valid for any numbers $z_a$.

By an appropriate choice of the numbers $z_a$ we can produce any transfer matrix on the left hand side of \eqref{Plucker}. Then the right hand side of \eqref{Plucker} will contain operators which are simpler in terms of the representation involved. Repeating this procedure, we can reduce all representations to the minimal ones. The right hand side will then only depend on the operators of the form ${\rm X}_{I_{k}}(z_0,\ldots,z_l)$ with $z_i-z_{i-1}=1$, which are exactly the {\rm Q}-functions. Interestingly, we conclude that any transfer matrix is a sum of $d$ terms, where $d$ is the dimension of the representation of the auxiliary space of the transfer matrix. 

The analytic properties of all operators present on the right hand side of \eqref{XdetQ} are fixed by \eqref{anal}. Furthermore, in line with our definition of Q-operators \eqref{Q-def}, the regular point of $\Top_{\Lambda}(z)$ is determined to be\footnote{For the convenience of the reader we stress that $\Top_{\Lambda}(z)$ in \eqref{EofT} corresponds to the transfer matrix built out of $\mathcal{R}_{\Lambda\Lambda}(z-\frac{n-1}{2})$. In particular, note that while $z_\star=\frac{n-1}{2}$, we have $z=0$ in \eqref{Hdensity}.} $z_\star=\frac{n-1}{2}$. The procedure presented above is rather complicated for generic representations, and we have not followed it through in detail. It is however straightforward to check by studying small representations that the energy formula is as expected given by the expression\footnote{For infinite-dimensional representations and rank one, a derivation in the framework of the algebraic Bethe ansatz can be found in \cite{Korchemsky:1994um}.}

\begin{equation}\label{energy.formula}
E=\sum_{i=1}^{n}\sum_{l=1}^{m_{I_{i}}}\left(\frac{1}{z^{I_i}_{l}-\sfrac{n-i-1}{2}-\ell_{n-i+1}-\sfrac{1}{2}}-\frac{1}{z^{I_i}_{l}-\sfrac{n-i-1}{2}-\ell_{n-i}+\sfrac{1}{2}}\right).
\end{equation}
A general proof is under construction.

\subsection*{Acknowledgements}
We would like to thank V.~Bazhanov, N.~Kanning, A.~Molev, V.~Mitev and Z.~Tsuboi for useful comments and correspondence. T.~{\L}ukowski is supported by a DFG grant in the framework of the SFB 647 {\it ``Raum - Zeit - Materie. Analytische und Geometrische Strukturen''}. C.~Meneghelli is partially supported by a DFG grant in the framework of the SFB 676 {\it ``Particles, Strings and the Early Universe"}.

\appendix
%%%%%%%%%%%%%%%%%%%%%%%%%%%%%%%%%%%%%%%%%%%%%%
\section{Review of Non-commutative Cayley-Hamilton Theorem}
\label{app:Reving}
%%%%%%%%%%%%%%%%%%%%%%%%%%%%%%%%%%%%%%%%%%

We present here a non-commutative version of the Cayley-Hamilton theorem following \cite{molevbook} (chapter~$7$).
It is used in  Section \ref{sec:oscpart} when solving the YBE for building blocks of Q-operators. It is remarkable that this theorem
is naturally embedded in the context of Yangians.

Let us first review the  well-known Cayley-Hamilton theorem for matrices.
Given a $q\times q$ matrix $M$ with commuting entries one has
\beq
p(M)=0\,,\qquad \text{where}\qquad\,\,\,\ p(z)\equiv\det(z-M)\,=\,z^q+a_{q}\,z^{q-1}+\dots+\,a_1\,,
\eeq
where $p(z)$ is called the characteristic polynomial.
The coefficients $a_i$ are symmetric functions of the eigenvalues of $M$.
The statement above can be equivalently rephrased saying that the matrix $M^q$ can be written as a linear combination of matrices $M^k$ with $k<q$ and coefficients $a_i$.

Imagine now that we have a matrix whose operatorial entries satisfy the $\alg{gl}(q)$ algebra.
Let us define the non-commutative analog of the characteristic polynomial known as a \emph{Capelli determinant}
\beq
\mathcal{C}(z)\equiv\,\sum_{\sigma \in S_q}\,(-1)^{\sigma}\,\left(z+J\right)^{\sigma(1)}_1 \dots\,\left(z+J-q+1 \right)^{\sigma(q)}_q\,.
\eeq 
We see that $\mathcal{C}(z)$ is by construction a polynomial in $z$ with coefficients in the universal enveloping algebra $\mathcal{U}(\alg{gl}(q))$ and can be written as
\beq
\label{preeelli}
\mathcal{C}(z)=\,z^q+\mathcal{C}_q\,z^{q-1}\,+\dots+\mathcal{C}_1\,.
\eeq
It can be proven  that the coefficients $\mathcal{C}_i$ belong to the center of  $\mathcal{U}(\alg{gl}(q))$. Moreover
\beq
\label{elli}
\mathcal{C}(z)=(z+\ell_1)\,\cdots\,(z+\ell_q)\,,
\eeq
where the Weyl group  acts on $\ell_i$ by permutations. The $\ell_i$ are, up to a global shift, the same as the shifted weights $\lambda_i'$. Again, $\mathcal{C}_i$ are symmetric functions of the $\ell$.
The non-commutative Cayley-Hamilton theorem states that
\beq
\label{noncommCH}
\mathcal{C}(-J^T+q-1)=0\,,\,\,\qquad\text{and}\qquad\,\,\mathcal{C}(-J)=0\,,
\eeq
where $T$ means transposition as explained in the following\footnote{For the convenience of the reader we point out that we will use a different looking, but equivalent use of transposition compared to \cite{molevbook}. This is done because we will use the second formula in \eqref{noncommCH}}. Powers of $J$ should be understood as, e.g.
\beq\label{powersofJ}
\left(J^{\,2}\right)^{A}_{B}=\,J^{C}_{B}\,J^{A}_{C}\,,\qquad\qquad A,B,C=1,\ldots ,q\,.
\eeq
This means contraction of indices from top-left to bottom-right and analogously for higher powers. On the other hand, by powers of $J^{\,T}$ we denote the other contraction
\beq
\left((J^{\,2})^{\,T}\right)^{A}_{B}=\,J^{A}_{C}\,J^{C}_{B}\,,\qquad\qquad A,B,C=1,\ldots ,q\,.
\eeq
Throughout this paper we use only the second formula of \eqref{noncommCH} together with \eqref{powersofJ}. As for the commutative case, \eqref{noncommCH} can be rewritten as
\beq
\label{noncommCHbis}
\left(J^{\,q}\right)^{A}_{B}=\,\sum_{k=0}^{q-1}\,\left(J^{\,k}\right)^{A}_{B}\,a_{k+1}(\ell_1,\dots,\ell_q)\,.
\eeq
We stress once more that $a_k(\ell_1,\dots,\ell_q)$ are symmetric functions of their arguments which can be written as
\beq\label{aofell}
a_{k}(\ell_1,\dots,\ell_q)=\,(-1)^{q+k}\,\sum_{j_0<j_1<\dots<j_{q-k}}\,\ell_{j_0}\,\dots\,\ell_{j_{q-k}}\,.
\eeq
We present here another formula we need in Section \ref{sec:oscpart}:
\begin{equation}\label{JofX}
(J^i)^{\dot a}_{a}=\sum_{k=1}^q (X^k)^{\dot a}_{a}\, \Delta_k(\hat \ell)\,(\hat{\ell}_k)^{i-1}\,,\qquad \Delta_k(\hat \ell)=\prod_{l\neq k}(\hat{\ell}_k-\hat{\ell}_l)^{-1}\,.
\end{equation}
%%%%%%%%%%%%%%%%%%%%%%%%%%%%%%%%%%%%%%%%%%%%%%
\section{Mathematica\texttrademark~Code for Q-operators}
\label{app:Qcode}
In this appendix\footnote{TL would like to thank Nils Kanning for inspiring discussions about the code.} we present a Mathematica code which allows to calculate Q-operators for compact representations. One has to specify the set $I=\{i_1,\ldots,i_p\}$ and the $\mathfrak{gl}(n)$ representation $\Lambda$ in the quantum space. The latter is of the form \texttt{$\Lambda$=$\{s_1,\ldots, s_n\}$} where $s_k$ is the number of boxes in the $k$-th row of the Young diagram. The code is grouped in three parts\footnote{In order to use the code it is sufficient to copy it, and to paste it into a Mathematica\texttrademark notebook.}: the first part calculates generators $J_{i_1}^{i_2}$ of the algebra in the representation $\Lambda$ with the function \texttt{JJ[$i_1$,$i_2$,$\Lambda$]}. We also define Casimir operators $c_i$ of the $\mathfrak{gl}(I)$ algebra with \texttt{CC[i,I,$\Lambda$]}. The second part defines the function \texttt{Q[L,z,I,$\Lambda$]}, which calculates the Q-operator $\Qop_{I}(z)$ for the spin chain of length $L$ and representation $\Lambda$ in the quantum space. Here $z$ is the spectral parameter. In the third part we present some functions  suitable for checking whether the Q-operators we constructed satisfy various basic properties such as proper commutation relations (using \texttt{comm[L,set1,set2,rep]}) and the QQ-relations (using \texttt{QQ[L,set,i1,i2,rep]}). 

\subsection{Generators of Algebras}
\texttt{ps[rep$\_$]:=ps[rep]=Block[$\{$res=(Prepend[rep,$\#$]$\&$/@(Tuples[Table[Range[rep[[\\
1,i+1]],rep[[1,i]]],$\{$i,1,Length[rep[[1]]]-1$\}$]]))$\}$,If[Length[res[[1,1]]]$==$1,\\
   Reverse@Sort@res,Reverse@Sort@Flatten[ps/@res,1]]];}\\
   \texttt{dim[rep$\_$]:=dim[rep]=Length[ps[$\{$rep$\}$]];}\\
   \texttt{J0[k$\_$][pat$\_$]:=J0[k][pat]=Sum[pat[[k,i]],$\{$i,1,k$\}$]-Sum[pat[[k-1,i]],$\{$i,1,k-1$\}$];}\\
  \texttt{Jp[k$\_$][pat$\_$]:=Jp[k][pat]=-Sum[Product[pat[[k,i]]-i-pat[[k+1,j]]+j,$\{$j,\\
  Range[k+1]$\}$]/Product[pat[[k,i]]-i-pat[[k,j]]+j,$\{$j,Complement[Range[k],$\{$i$\}$]$\}$]$\\
   v[$ReplacePart[pat,$\{$k,i$\} $->$ $pat[[k,i]]+1]$] $,$\{$i,1,k$\}$];}\\
   \texttt{Jm[k$\_$][pat$\_$]:=Jm[k][pat]=Sum[Product[pat[[k,i]]-i-pat[[k-1,j]]+j,$\{$j,\\
   Range[k-1]$\}$]/Product[pat[[k,i]]-i-pat[[k,j]]+j,$\{$j,Complement[Range[k],$\{$i$\}$]$\}$]$\\
   v[$ReplacePart[pat,$\{$k,i$\} $->$ $pat[[k,i]]-1]$] $,$\{$i,1,k$\}$];}\\
   \texttt{JJ[k$\_$,l$\_$,rep$\_$]:=JJ[k,l,rep]=Which[k$==$l,DiagonalMatrix[J0[k]/@ps[$\{$rep$\}$]],\\
   l$==$k+1,Transpose[Table[Coefficient[Jp[k][ps[$\{$rep$\}$][[i]]],$v[\#]$]$\&$/@\\
   ps[$\{$rep$\}$],$\{$i,1,Length[ps[$\{$rep$\}$]]$\}$]],l$==$k-1,Transpose[Table[Coefficient[\\
   Jm[l][ps[$\{$rep$\}$][[i]]],$v[\#]
   $]$\&$/@ps[$\{$rep$\}$],$\{$i,1,Length[ps[$\{$rep$\}$]]$\}$]],l$>$k+1,\\
   JJ[k,k+1,rep].JJ[k+1,l,rep]-JJ[k+1,l,rep].JJ[k,k+1,rep],l$<$k-1,JJ[k,k-1,rep].\\
   JJ[k-1,l,rep]-JJ[k-1,l,rep].JJ[k,k-1,rep]];}\\
   \texttt{CC[k$\_$,set$\_$,rep$\_$]:=CC[k,set,rep]=Sum[Dot[Sequence@@Table[JJ[j[i+1],j[i],rep],\\
   $\{$i,1,k$\}$]]/.j[k+1]$ $->$ $j[1],Evaluate[Sequence@@Array[$\{$j[$\#$],set$\}\&$,k]]];}\\
   \texttt{CCDiag[k$\_$,set$\_$,rep$\_$]:=CCDiag[k,set,rep]=Diagonal[Inverse[Transpose[\\
   Eigenvectors[CC[Length[set],set,rep]]]].CC[k,set,rep].Transpose[Eigenvectors[\\
   CC[Length[set],set,rep]]]]}
   
   An example: the second Casimir of the $\mathfrak{gl}(2)$ subalgebra of $\mathfrak{gl}(4)$ generated by the set $\{1,3\}$.  The representation of $\mathfrak{gl}(4)$ is taken to be the fundamental representation \texttt{rep}=$\{1,0,0,0\}$.\\
 \texttt{CC[2,\{1,3\},\{1,0,0,0\}]} 
    
   \subsection{Q-operators}
  \texttt{CasimirOfL[p$\_$,k$\_$]:=Sum[Product[(1+1/(l[i]-l[j])),$\{$j,Complement[Range[p],$\{$i$\}$]$\}$]\\
  Power[l[i],k],$\{$i,1,p$\}$];}\\
   \texttt{CHS[z$\_$,set$\_$,rep$\_$]:=Block[$\{$n=Length[rep],p=Length[set]$\}$,If[p$>$0,(Transpose[\\
   Eigenvectors[CC[p,set,rep]]].DiagonalMatrix[Product[(Gamma[z-l[i1]-(p)/2+1]\\
   /.(Table[(NSolve[Table[CasimirOfL[p,i2]$==$CCDiag[i2,set,rep][[i4]],$\{$i2,1,p$\}$],\\
   Table[l[i3],$\{$i3,1,p$\}$]][[1]]),$\{$i4,1,Length[CC[1,set,rep]]$\}$]//Rationalize)),\\
   $\{$i1,1,p$\}$]/Product[Gamma[z-(rep[[i5]]-i5+1)-p/2+1],$\{$i5,1,n$\}$]Product[\\
   Gamma[z-rep[[n]]-(-i6+1)-p/2+1],$\{$i6,p+1,n$\}$]//FunctionExpand].Inverse[Transpose[\\
   Eigenvectors[CC[p,set,rep]]]]//Expand),Product[Gamma[z-rep[[n]]-(-i1+1)+1]/\\
   Gamma[z-(rep[[i1]]-i1+1)+1],$\{$i1,1,n$\}$]IdentityMatrix[dim[rep]]//FunctionExpand]];}\\
  \texttt{expp[set$\_$,rep$\_$][A$\_$]:=Block[$\{$n=Length[rep],m=rep[[1]],t$\}$,t=Flatten[Tuples[Tuples[\\
  $\{$Complement[Range[n],set],set$\}$],$\#$]$\&$/@Range[1,m],1];Total[KroneckerProduct@@@($\{$\\
  Dot@@(JJ[Sequence@@$\#$,rep]$\&$/@$\#$),Power[-1,Length[$\#$]]/Factorial[Length[$\#$]]D[A,\\
  Sequence@@$\#$]$\&$@(x[Sequence@@Reverse[$\#$]]$\&$/@$\#$)$\}\&$/@t)]+KroneckerProduct[\\
  IdentityMatrix[dim[rep]],A]];}\\
  \texttt{expl[set$\_$,rep$\_$]:=Block[$\{$n=Length[rep],m=rep[[1]],t$\}$,t=Tuples[$\{$set,Complement[\\
  Range[n],set]$\}$];Dot@@((IdentityMatrix[dim[rep]]+Sum[Power[x[Sequence@@$\#$],i]/i!\\
  Dot[Sequence@@ConstantArray[JJ[Sequence@@$\#$,rep],i]],$\{$i,1,m$\}$])$\&$/@t)];}\\
   \texttt{Lax[z$\_$,set$\_$,rep$\_$][A$\_$]:=Block[$\{$n=Length[rep],p=Length[set]$\}$,If[0$<$p$<$n,((\\
   KroneckerProduct[expl[set,rep].(CHS[z,Complement[Range[n],set],rep]//N//\\
   ExpandAll//Chop//Rationalize//FunctionExpand),IdentityMatrix[Length[A]]]//\\
   ArrayFlatten).(expp[set,rep][A])//FunctionExpand//Simplify//Expand),CHS[z,\\
   Complement[Range[n],set],rep]]];}\\
   \texttt{Q[L$\_$,z$\_$,set$\_$,rep$\_$]:=Block[$\{$a,p=Length[set],n=Length[rep]$\}$,If[0$<$p$<$n,a=(Nest[\\
   ArrayFlatten[Lax[z,set,rep][$\#$]]$\&$,$\{\{$Product[Power[x[i1,i2],nn[i1,i2]],$\{$i1,set$\}$,\\
   $\{$i2,Complement[Range[Length[rep]],set]$\}$]$\}\}$,L]/Product[Power[x[i1,i2],nn[i1,i2]],\\
   $\{$i1,set$\}$,$\{$i2,Complement[Range[Length[rep]],set]$\}$]//Expand)/.Power[x[i1$\_$,i2$\_$],i$\_$.\\
  ] :$>$0//Expand,a=If[L$>$1,KroneckerProduct[Sequence@@ConstantArray[Lax[z,set,rep\\
  ][ $\{$1$\}$],L]],Lax[z,set,rep][$\{$1$\}$]]];Exp[I (z-Last[rep])Sum[$\phi $[i],$\{$i,set$\}$]]a/.$\{$\\
   Power[nn[j$\_$,i$\_$],k$\_$.]:$>$ -(-1+Exp[I ($\phi$[i]-$\phi $[j])]) HurwitzLerchPhi[\\
   Exp[I ($\phi $[i]-$\phi $[j])],-k,0]$\}$]}
   
   An example: Q-operator $\Qop_{\{1\}}(z)$ for the length $L=2$ spin chain with the spin-1 representation (\texttt{rep}=$\{2,0\}$) in the quantum space. \\
   \texttt{Q[2,z,\{1\},\{2,0\}]//Simplify}
   
   \subsection{Checks}
\texttt{comm[L$\_$,set1$\_$,set2$\_$,rep$\_$]:=Q[L,z1,set1,rep].Q[L,z2,set2,rep]-Q[L,z2,set2,rep].\\
Q[L,z1,set1,rep]//Simplify;}\\
\texttt{QQ[L$\_$,set$\_$,i1$\_$,i2$\_$,rep$\_$]:=Q[L,z+1/2,Union[set,$\{$i1$\}$],rep].Q[L,z-1/2,Union[set,\\
$\{$i2$\}$],rep]-Q[L,z-1/2,Union[set,$\{$i1$\}$],rep].Q[L,z+1/2,Union[set,$\{$i2$\}$],rep]-2 i \\
Sin[1/2 ($\phi $[i1]-$\phi$[i2])]Q[L,z,Union[set,$\{$i1$\}$,$\{$i2$\}$],rep].Q[L,z,set,rep]//Simplify;}

An example: Commutation relation between $\Qop_{1}(z_1)$ and $\Qop_2(z_2)$ for the length $L=2$ spin chain in the adjoint representation of $\mathfrak{gl}(3)$ (\texttt{rep}=$\{2,1,0\}$).\\
\texttt{comm[2,\{1\},\{2\},\{2,1,0\}]}

  An example: QQ-relation for the antifundamental $\mathfrak{gl}(3)$ representation  \texttt{rep}=$\{1,1,0\}$. We take the length $L=2$ and the plaquette containing operators $\Qop_{\{1\}}(z)$, $\Qop_{\{1,2\}}(z)$, $\Qop_{\{1,3\}}(z)$ and $\Qop_{\{1,2,3\}}(z)$.\\
  \texttt{QQ[2,\{1\},2,3,\{1,1,0\}]}
%%%%%%%%%%%%%%%%%%%%%%%%%%%%%%%%%%%%%%%%%%%%%%%%%%%%%%%%%
\section{Comparing R-matrix and Tensor Product Graph Method}
\label{app:TPG}
%%%%%%%%%%%%%%%%%%%%%%%%%%%%%%%%%%%%%%%%%%%%%%%%%%%%%%%%%

We want to prove that the formula \eqref{Rllsolution} which we found directly from the YBE reproduces the one found using the tensor product graph method (see e.g.~\cite{Gould:2002de}). Let us take any rectangular representation space $V$. Then the decomposition of the tensor product of $V$ is multiplicity-free and reads  
\begin{equation}\label{decomposition}
V\otimes V=\oplus V_{\lambda}\,.
\end{equation}
We introduce a relation $\sim$ for representations appearing in the decomposition \eqref{decomposition} as
\begin{equation}
V_{\lambda}\sim V_{\mu} \Longleftrightarrow V_\lambda \subset V_{adj}\otimes V_{\mu} \mbox{   and $V_{\lambda}$ has different parity than $V_{\mu}$}\,.
\end{equation}
The relation $\sim$ is symmetric and one can show that
\begin{equation}\label{TPGdef}
V_{\lambda}\sim V_{\mu} \Longleftrightarrow \sum_{l=1}^n |A_{l}^{\lambda}-A_{l}^{\mu}|=2\,.
\end{equation}
In words, it means that two representations are in the relation $\sim$ if the former can be obtained from the latter by moving a single box in the Young tableaux.

From the tensor product graph method we have
\begin{equation}\label{TPG}
\check R_{\Lambda\Lambda}(z)=\sum_{\lambda}\rho_{\lambda}(z)P_{\lambda}\,,
\end{equation}  
with
\begin{equation}
\frac{\rho_{\lambda}(z)}{\rho_{\mu}(z)}=\frac{\delta-z}{\delta+z}\, ,\qquad \delta=\sfrac{1}{4}(C_2(\lambda)-C_2(\mu))\,\, \mbox{  if only  }\lambda \sim\mu\,. 
\end{equation} 
Let us calculate the second Casimir
\begin{equation}
C_2=\sum_{i=1}^{2a}\prod_{j\neq i}\left( 1-\frac{1}{\hat{A}_i-\hat{A}_j}\right)\hat{A}_i^2=\sum_{i=1}^{2a}\left( \hat{A}_i^2-(2a-1)\hat{A}_i\right)+\frac{(2a-2)(2a-1)2a}{6}\,.
\end{equation}
Let us take $\lambda, \mu$ such that $\lambda \sim\mu$. Then using \eqref{TPGdef} we have 
\begin{equation}
C_2(\lambda)-C_2(\mu)=\hat{A}_k^2+\hat{A}_{\bar k}^2-(\hat{A}_k+1)^2-(\hat{A}_{\bar k}^2-1)^2=-2\hat{A}_k+2\hat{A}_{\bar k}-2
\end{equation}
and
\begin{equation}
\frac{\rho_{\lambda}(z)}{\rho_{\mu}(z)}=\frac{\sfrac{1}{2}(\hat{A}_{\bar k}-\hat{A}_k-1)-z}{\sfrac{1}{2}(\hat{A}_{\bar k}-\hat{A}_k-1)+z}\,.
\end{equation}
On the other hand from our formula we have
\begin{eqnarray}
\frac{R_{\lambda}(z)}{R_{\mu}(z)}&=& \frac{\Gamma\left( z+\sfrac{1}{2}(\hat{A}_k-\hat{A}_{\bar k}+1)\right)\Gamma\left( z+\sfrac{1}{2}(\hat{A}_{\bar k}-\hat{A}_k+1)\right)}{\Gamma\left( z+\sfrac{1}{2}(\hat{A}_k-\hat{A}_{\bar k}+1)+1\right)\Gamma\left( z+\sfrac{1}{2}(\hat{A}_{\bar k}-\hat{A}_k+1)-1\right)}\\
&=& \frac{z+\sfrac{1}{2}(\hat{A}_k-\hat{A}_{\bar k}+1)}{z+\sfrac{1}{2}(\hat{A}_{\bar k}-\hat{A}_k-1)}=- \frac{\sfrac{1}{2}(\hat{A}_{\bar k}-\hat{A}_k+1)-z}{\sfrac{1}{2}(\hat{A}_{\bar k}-\hat{A}_k-1)+z}\,.
\end{eqnarray}
The overall minus sign comes from the fact that in the tensor product graph method we considered $\check R(z)=P R(z)$. From the definition of the relation $\sim$ we know that representations $\lambda$ and $\mu$ have different parities then we can conclude that
\begin{equation}
\frac{\check R_{\lambda}(z)}{\check R_{\mu}(z)}=\frac{\rho_{\lambda}(z)}{\rho_{\mu}(z)}\,,
\end{equation}
proving the equivalence of the results \eqref{Rllsolution} and \eqref{TPG}.
%%%%%%%%%%%%%%%%%%%%%%%%%%%%%%%%%%%%%%%%%%

%%%%%%%%%%%%%%%%%%%%%%%%%%%%%%%%%%%%%%%%%%%%%%%%%%%%%%%%%%%%%%%%%%%%
\section{Derivation of Hamiltonian from R-matrix}
\label{app:ham}
%%%%%%%%%%%%%%%%%%%%%%%%%%%%%%%%%%%%%%%%%%%%%%%%%%%%%%%%%
We use the formula \eqref{Hdensity} together with \eqref{Rllsolution} and the normalization \eqref{Rllnorm} in order to find an expression for the Hamiltonian density
\begin{eqnarray}\label{app.ham1}
\mathcal{H}&=&-\frac{d}{dz}\log\Rlam(z)\Big| _{z=0}\\\label{app.ham2}
&=&- \sum_{k=1}^a\left[ \psi\left(z+\frac{\hat{A}_k-\hat{A}_{\bar k}+1}{2}\right)+ \psi\left(z+\frac{\hat{A}_{\bar k}-\hat{A}_k+1}{2}\right)\right.\\\nonumber
&&\left.- \psi\left(z+\frac{\hat{A}_k+\hat{A}_{\bar k}+1}{2}+2a-k\right)- \psi\left(z-\frac{\hat{A}_k+\hat{A}_{\bar k}+1}{2}-2a+k+1\right)\right]\Big|_{z=0}\,.
\end{eqnarray}
One has to proceed carefully because some of terms in \eqref{app.ham2} are divergent while evaluated at $z=0$. Still the result is finite because of the relation
\begin{equation} 
\lim_{\epsilon\to 0}\left[\psi(\epsilon-x)-\psi(\epsilon-y)\right]=\psi(x+1)-\psi(y+1)\,, \qquad \mbox{for }x,y\in \mathbb{N}
\end{equation}
Using the fact that for eigenvalues of $\hat{A}$ we have
\begin{equation}
\frac{A_{\bar k}-A_k+1}{2}\in \mathbb{Z}_{-}\,, \qquad -\frac{A_k+A_{\bar k}+1}{2}-2a+k+1\in \mathbb{Z}_{-}\,,
\end{equation}
we can write
\begin{eqnarray}
\mathcal{H}&=&-2 \sum_{k=1}^a\left[ \psi\left(\frac{\hat{A}_k-\hat{A}_{\bar k}+1}{2}\right)- \psi\left(\frac{\hat{A}_{\bar k}+\hat{A}_k+1}{2}+2a-k\right)\right]\\
&=& -2 \sum_{k=1}^a\left[ \psi\left(\frac{\hat{A}_k-\hat{A}_{\bar k}+1}{2}\right)- \psi\left(s+a-k+1\right)\right],
\end{eqnarray}
where the last relation holds true for the compact representations with $a$ rows and $s$ columns. 

%%%%%%%%%%%%%%%%%%%  BIBLIOGRAPHY  %%%%%%%%%%%%%%%%%%%%%%%%%%%%%%%%%%%%%%

\bibliographystyle{utphys}
\bibliography{flms}

\end{document}